\documentclass[fleqn,usenatbib]{mnras}

\usepackage{newtxtext,newtxmath}

\usepackage[T1]{fontenc}

\DeclareRobustCommand{\VAN}[3]{#2}
\let\VANthebibliography\thebibliography
\def\thebibliography{\DeclareRobustCommand{\VAN}[3]{##3}\VANthebibliography}

\usepackage{graphicx}	
\usepackage{amsmath}	
\usepackage{xcolor}
\usepackage{derivative}
\usepackage{bm,url}
\newcommand{\st}{\operatorname{St}}

\defcitealias{Smallwood2024}{Paper I}

\title[Polar dusty circumbinary discs]{Polar alignment of a dusty circumbinary disc -- II. Application to 99 Herculis}

\author[Smallwood et al.]{Jeremy L. Smallwood,$^{1}$\thanks{E-mail: jlsmallwood@asiaa.sinica.edu.tw}
Min-Kai Lin,$^{1,2}$
Rebecca Nealon,$^{3,4}$
Hossam Aly$^{5,6}$
and
Cristiano Longarini$^{7,8}$
\\
$^{1}$Institute of Astronomy and Astrophysics, Academia Sinica, Taipei 106216, R.O.C.\\
$^{2}${Physics Division, National Center for Theoretical Sciences, Taipei 106216, Taiwan}\\
$^{3}$Centre for Exoplanets and Habitability, University of Warwick, Coventry CV4 7AL, UK\\
$^{4}$Department of Physics, University of Warwick, Coventry CV4 7AL, UK\\
$^{5}$Faculty of Aerospace Engineering, Delft University of Technology, Kluyverweg 1, 2629 HS Delft, The Netherlands\\
${6}$ Zentrum fur Astronomie der Universit ¨ at Heidelberg, Astronomisches Rechen-Institut, M'onchhofstr 12-14, D-69120 Heidelberg, Germany \\
$^{7}$Dipartimento di Fisica, Università degli Studi di Milano, via Celoria 16, 20133 Milano, Italy\\
$^{8}$Institute of Astronomy, University of Cambridge, Madingley Rd, CB3 0HA Cambridge, United Kingdom\\
}

\date{Accepted XXX. Received YYY; in original form ZZZ}

\pubyear{2024}

\begin{document}
\label{firstpage}
\pagerange{\pageref{firstpage}--\pageref{lastpage}}
\maketitle

\begin{abstract}
We investigate the formation of dust traffic jams in polar-aligning circumbinary discs. In our first paper, we found as the circumbinary disc evolves towards a polar configuration perpendicular to the binary orbital plane, the differential precession between the gas and dust components leads to multiple dust traffic jams. These dust traffic jams evolve to form a coherent dust ring. In part two, we use 3D smoothed particle hydrodynamical simulations of gas and dust to model an initially highly misaligned circumbinary disc around the 99 Herculis (99 Her) binary system.   Our results reveal that the formation of these dust rings is observed across various disc parameters, including the disc aspect ratio, viscosity, surface density power law index, and temperature power law index. The dust traffic jams are long-lived and persist even when the disc is fully aligned polar. The midplane dust-to-gas ratio within the rings can surpass unity, which may be a favourable environment for planetesimal formation. Using 2D inviscid shearing box calculations with parameters from our 3D simulations, we find streaming instability modes with significant growth rates. The streaming instability growth timescale is less than the tilt oscillation timescale during the alignment process. Therefore, the dust ring will survive once the gas disc aligns polar, suggesting that the streaming instability may aid in forming polar planets around 99 Her.
\end{abstract}




\section{Introduction}
In our galaxy, it is estimated that a significant portion, ranging from 40 per cent to 50 per cent of stars exist as part of binary systems \citep{Duquennoy1991,Mayor2011,Raghavan2010,Tokovinin2014a,Tokovinin2014b}. Young binary star systems often exhibit a circumbinary disc composed of gas and dust, which serve as potential sites for planet formation. Observations have consistently shown that circumbinary discs commonly exhibit misalignment with respect to the binary orbital plane \cite[e.g.,][]{Czekala2019}. When an initially misaligned circumbinary disc surrounds a binary system with a circular orbit, it gradually aligns itself to the binary orbital plane \cite[e.g.,][]{papaloizou1995,Lubow2000,Nixon2011,Facchini2013,Foucart2014}. However, if the binary system possesses a non-zero eccentricity, a low-mass circumbinary disc with a substantial misalignment will precess around the binary eccentricity vector. This precession causes the angular momentum vector of the disc to align with the binary eccentricity vector, resulting in a polar-aligned circumbinary disc \citep{Aly2015,Martinlubow2017,Martin2018,Lubow2018,Zanazzi2018}. In the case of a massive disc, alignment to a generalized polar state occurs with a lower degree of misalignment to the binary orbital plane \citep{Zanazzi2018,MartinLubow2019}.

The impact of this misalignment on the dynamics of gas, dust, and the formation of circumbinary planets is not yet fully understood. The presence of a binary star system exerts a torque on the disc, potentially influencing the planet formation process compared to discs around single stars \citep{Nelson2000,Mayer2005,Boss2006,Martin2014,Fu2015a,Fu2015b,Fu2017}. By gaining a deeper understanding of the structure and evolution of dusty circumbinary discs, we can gain insights into the characteristics of exoplanets.

\cite{Smallwood2024} (\citetalias{Smallwood2024}) focused on investigating the formation of dust traffic jams within generic circumbinary discs undergoing polar alignment. Initially, we utilized a 1D model developed by \cite{Aly2021} to demonstrate the formation of dust rings in a misaligned circumbinary disc during the process of polar alignment. Subsequently, we conducted Smoothed Particle Hydrodynamics (SPH) simulations incorporating gas and dust components to model an initially highly misaligned circumbinary disc around an eccentric binary system. As the circumbinary disc gradually transitioned to a polar state, the differential precession between the gas and dust components led to the formation of dust traffic jams, which ultimately evolved into dense dust rings. The 1D and 3D approaches yielded similar results. Furthermore, our findings revealed that the formation of these dust rings was influenced by factors such as Stokes number, binary eccentricity, and the initial tilt of the disc. Once the disc reaches a polar alignment, the mechanism responsible for generating dust rings ceases. However, if the dust rings were formed before the polar alignment, they could persist in the polar state. The existence of stable polar dust rings bears implications for the formation of polar planets. We apply our results to a well-known binary star system, 99 Herculis hosting a circumbinary debris disc.

99 Herculis (99 Her) is a binary star system consisting of an  F7V primary star and a k4V secondary star, with an estimated age of $6-10\, \rm Gyr$ \citep{Nordstrom2004,Takeda2007} at a distance of $15.64\, \rm pc$ \citep{vanLeeuwen2008}. Observations of the binary system date back to the latter half of the 1800s \cite[e.g.,][]{Burnham1878,Flammarion1879,Gore1890}. In recent years,  higher resolution observations of 99 Her have resulted in a more precise measurement of the binary orbital parameters \citep{Kennedy2012}. The binary separation, eccentricity, inclination (with respect to the sky plane), and orbital period are $a = 16.5\, \rm au$, $e_{\rm b} = 0.766$, $i = 39^\circ$, and $P_{\rm orb} = 56.3\, \rm yr$. The longitude of the ascending node is $\Omega = 41^\circ$. The longitude of the pericentre  is measured anticlockwise from the ascending node, projected onto the sky plane, which has a position angle (PA) of $163^\circ$, resulting in a value of $\omega = 116^\circ$. Since the binary is inclined, the sum of $\Omega$ and $\omega$ do not equal the sky plane PA. The mass of the primary star is $M_1 = 0.94\, \rm M_{\odot}$ and the mass of the secondary star is $M_2 = 0.46\, \rm M_{\odot}$, giving a total mass of $M = M_1 + M_2 = 1.4\, \rm M_{\odot}$. 

The 99 Her binary system hosts a circumbinary debris disc that was first detected using the {\it Herschel} Photodetector and Array Camera and Spectrometer \cite[PACS;][]{Griffin2010,Poglitsch2010}. Using resolved PACS images of the debris disc, \cite{Kennedy2012} estimated the debris disc structure, inclination, and PA using two-dimensional Gaussian models. To fit the observational data, their debris ring model was located in a small range of radii near $120\, \rm au$. Due to disc emission wavelengths being measured insufficiently, the grain properties and size distributions could not be accurately constrained. The debris disc has an observed PA of $72^\circ$. Given that the projection of the binary pericentre direction of the sky has a PA of $163^\circ \pm 2^\circ$ and a vector normal to this has a PA of $73^\circ \pm 2^\circ$, the debris disc is tilted $87^\circ$ with respect to the binary pericentre direction \citep{Kennedy2012}. The observed disc tilt is thus only $3^\circ$ away from polar alignment. Another polar disc has been observed around HD 98800B but shows traces of gas still present \citep{Kennedy2019}.

In this work,  we further the work of our \citetalias{Smallwood2024} by investigating polar-aligning dusty circumbinary discs with a comprehensive comparison of hydrodynamical simulations. We apply the results to 99 Herculis, which hosts the only known polar circumbinary debris disc. This endeavor is aimed at enhancing our comprehension of the formation and evolution of dust ring structures within the context of varying disc aspect ratio, surface density, temperature, and viscosity parameters. 
By conducting this comparative analysis, we will uncover intricate patterns, trends, and relationships between these disc parameters and the resulting characteristics of the dust ring structures in initially misaligned circumbinary discs. 



The paper is organized as follows. Section~\ref{sec::methods} describes the setup for our hydrodynamical simulations of a misaligned gaseous and dusty circumbinary disc.  In Section~\ref{sec::hydro_results}, we report the results of our hydrodynamical simulations. In Section~\ref{sec::Streaming_instability}, we estimate analytically the streaming instability growth rate given our hydro results.  Lastly, we give a summary in Section~\ref{sec:Summary}.

\section{Methods}
\label{sec::methods}
To simulate an inclined dusty circumbinary disc around 99 Herculis, we utilize the three-dimensional smoothed particle hydrodynamics code {\sc phantom} \citep{Price2018}. The code employs different techniques for modeling dust-gas mixtures based on the Stokes number of the dust grains. In the two-fluid implementation, dust and gas are treated as separate discrete particles, considering drag forces and explicit time-stepping \citep{Laibe2012a,Laibe2012b}. Conversely, in the one-fluid version, dust is treated as part of the overall mixture, and its evolution equations account for the dust fraction \citep{Price2015a}.  We adopt the two-fluid algorithm in our specific case of investigating the polar alignment of a circumbinary disc with low gas density and St greater than 1, where $\rm St$ is the Stokes number of the dust grain. The Stokes number is given by the ratio of the particle's stopping time to the local dynamical time scale of the gas in the disc. The two-fluid implementation accounts for drag heating but does not consider the thermal coupling between the dust and gas \cite[refer to][for more details]{Laibe2012a}. Table~\ref{table::setup} summarises the suite of hydrodynamical simulations.

\begin{table}
	\centering
	\caption{The setup of the gas and dust SPH simulations that includes an initial circumbinary disc. The table lists the initial disc aspect ratio, $H/r$ at $r_{\rm in}$, viscosity parameter, $\alpha_{\rm SS}$, surface density power law index, $p$, and the sound speed power law index, $q$. Note that the temperature power law index is given as $2q$. }
	\begin{tabular}{lcccc} 
		\hline
	    Model & $H/r$  & $p$ & $q$ & $\alpha_{\rm SS}$\\
		\hline
		\hline
		run1 & $0.1$   & $+3/2$ & $+3/4$ & $0.01$ \\
            run2 & $0.05$   & $+3/2$ & $+3/4$ & $0.01$ \\
            run3 & $0.1$   & $+1$ & $+3/4$ & $0.01$ \\
		run4 & $0.1$   & $+3/2$ & $+1/2$ & $0.01$ \\
            run5 & $0.1$   & $+3/2$ & $+3/4$ & $0.005$ \\
  	    run6 & $0.1$   & $+3/2$ & $+3/4$ & $0.02$ \\
		\hline
	\end{tabular}
    \label{table::setup}
\end{table}

\begin{figure*} 
\centering
\includegraphics[width=0.34\columnwidth]{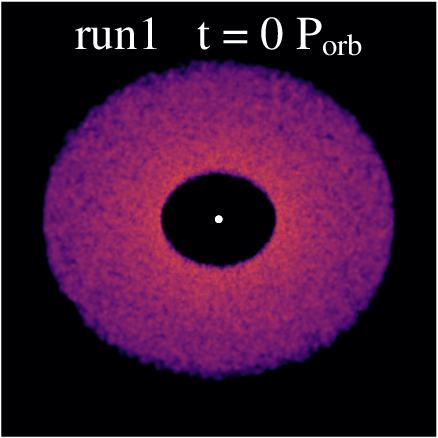}
\includegraphics[width=0.34\columnwidth]{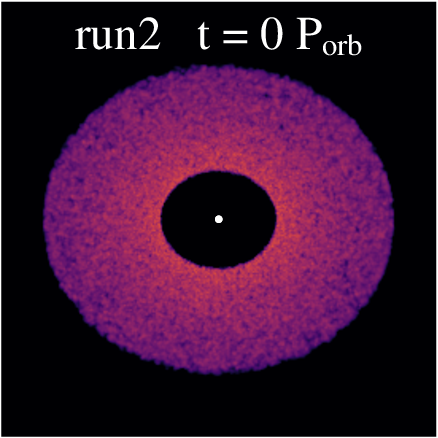}
\includegraphics[width=0.34\columnwidth]{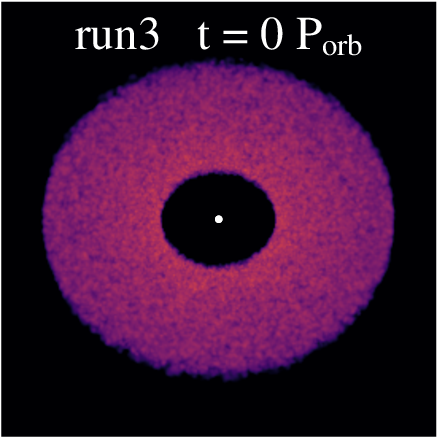}
\includegraphics[width=0.34\columnwidth]{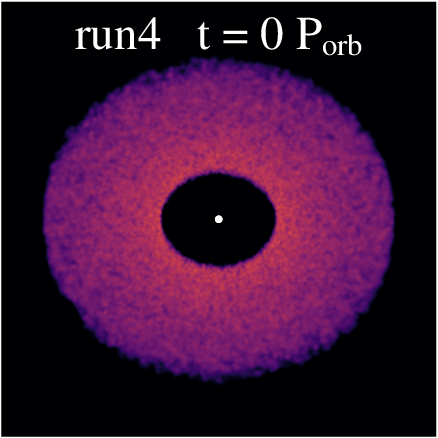}
\includegraphics[width=0.34\columnwidth]{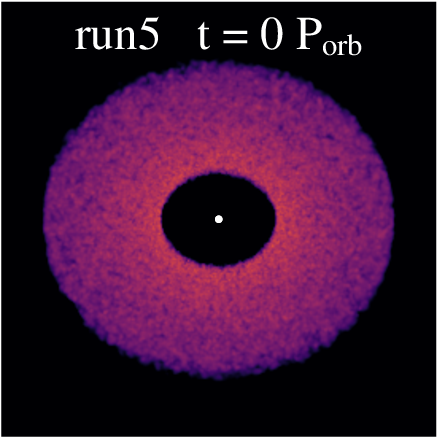}
\includegraphics[width=0.34\columnwidth]{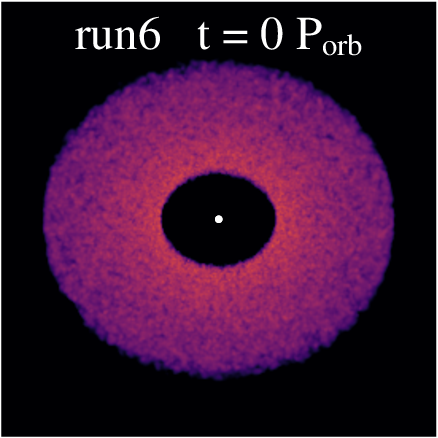}
\includegraphics[width=0.34\columnwidth]{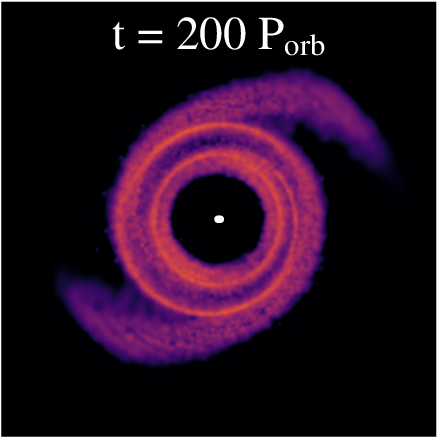}
\includegraphics[width=0.34\columnwidth]{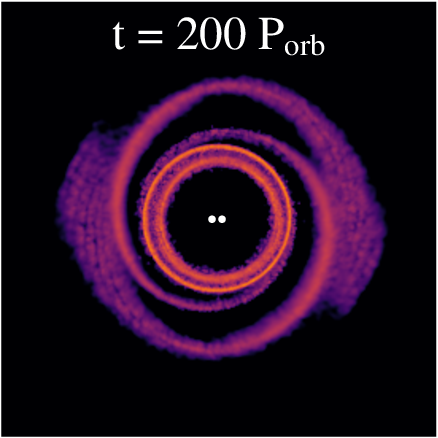}
\includegraphics[width=0.34\columnwidth]{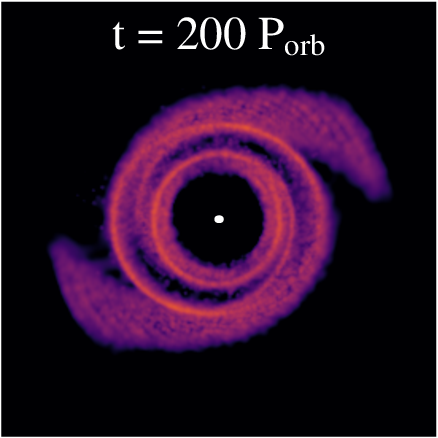}
\includegraphics[width=0.34\columnwidth]{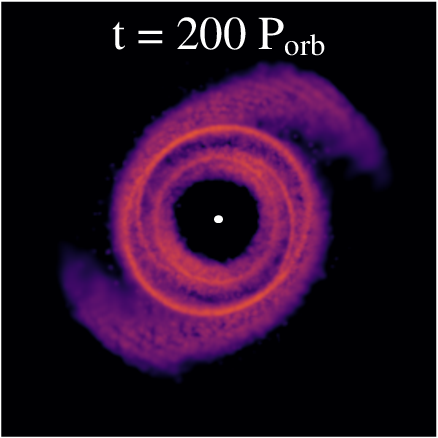}
\includegraphics[width=0.34\columnwidth]{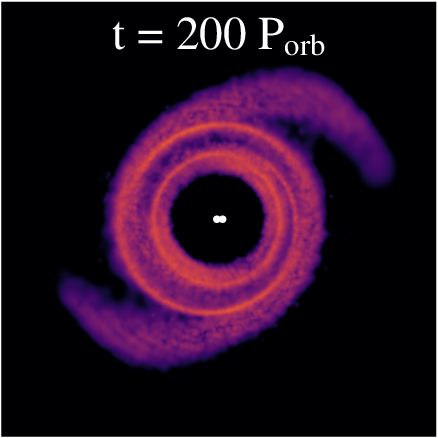}
\includegraphics[width=0.34\columnwidth]{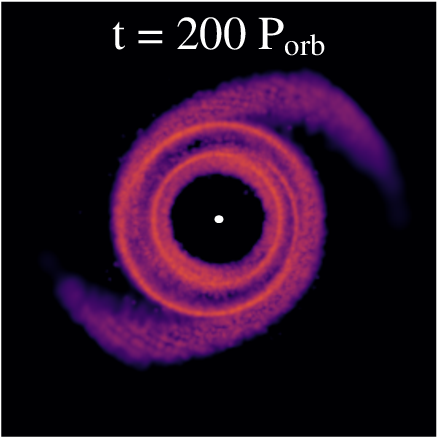}
\includegraphics[width=0.34\columnwidth]{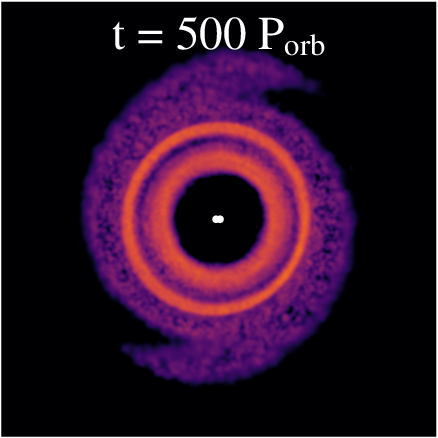}
\includegraphics[width=0.34\columnwidth]{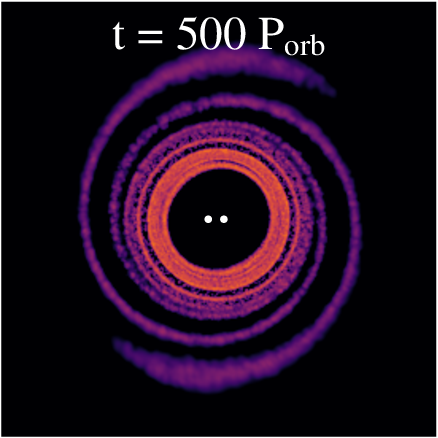}
\includegraphics[width=0.34\columnwidth]{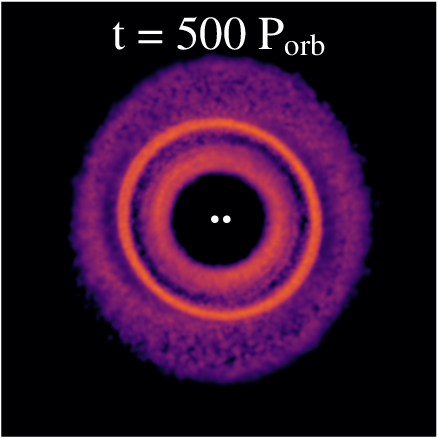}
\includegraphics[width=0.34\columnwidth]{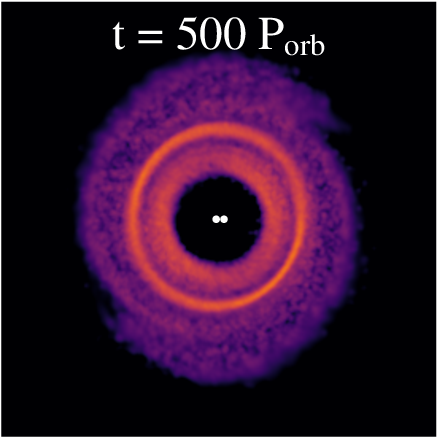}
\includegraphics[width=0.34\columnwidth]{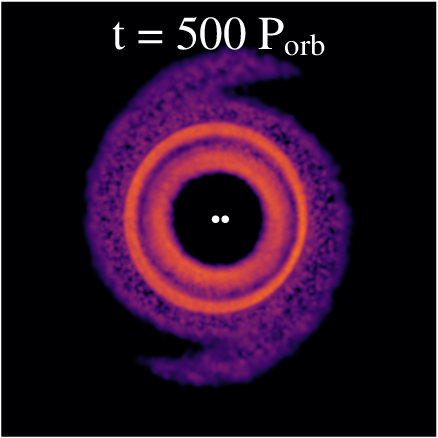}
\includegraphics[width=0.34\columnwidth]{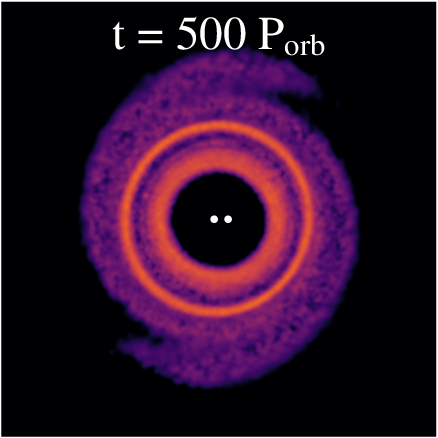}
\includegraphics[width=0.34\columnwidth]{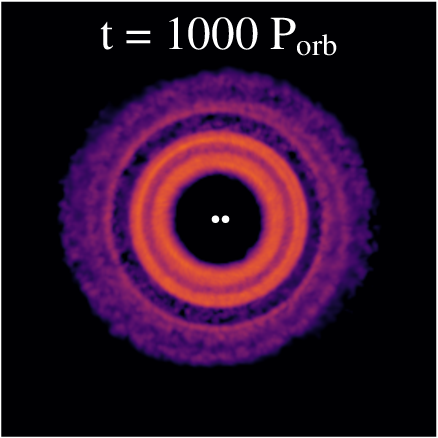}
\includegraphics[width=0.34\columnwidth]{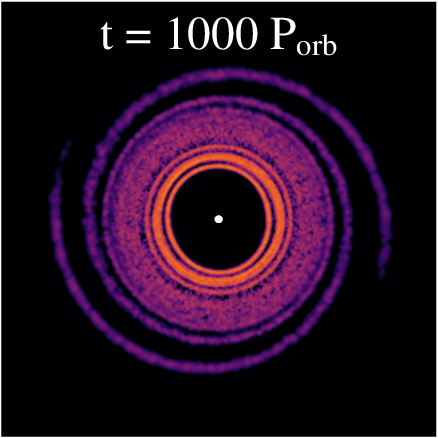}
\includegraphics[width=0.34\columnwidth]{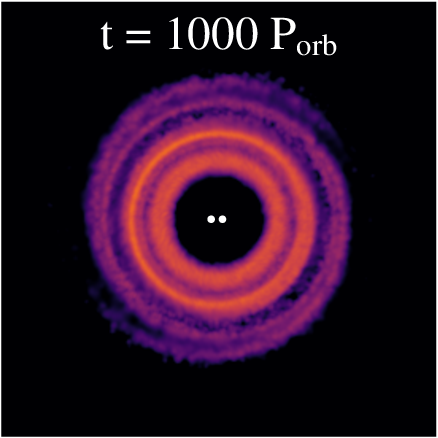}
\includegraphics[width=0.34\columnwidth]{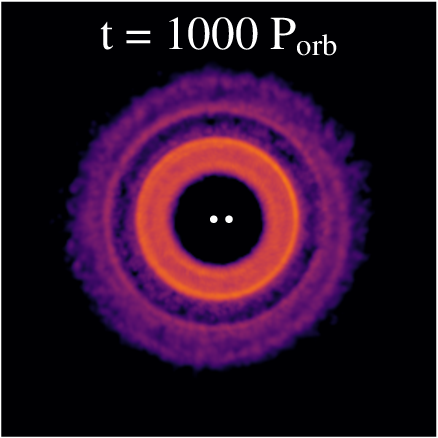}
\includegraphics[width=0.34\columnwidth]{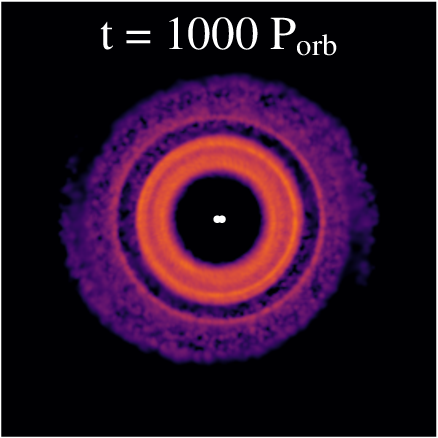}
\includegraphics[width=0.34\columnwidth]{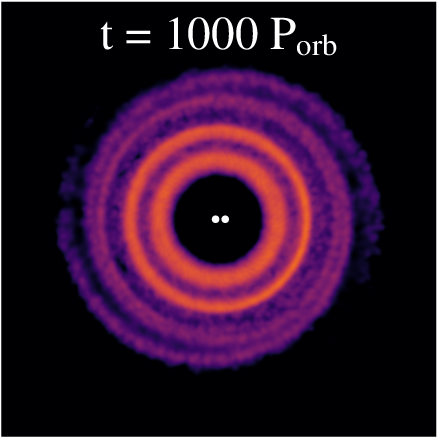}
\includegraphics[width=2.084\columnwidth]{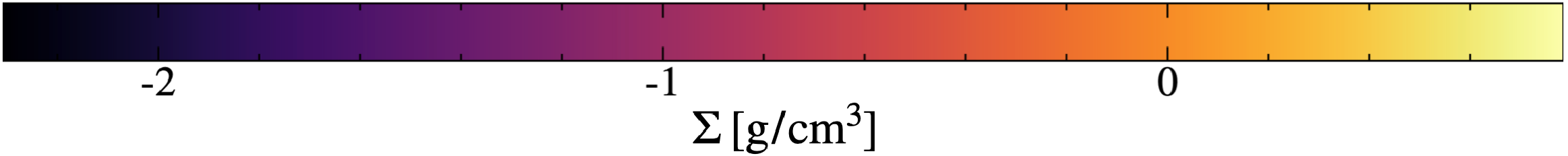}
\caption{The evolution of the initially misaligned dusty circumbinary disc around 99 Her for each hydrodynamical simulation, run1 (first column), run2 (second column), run3 (third column), run4 (fourth column), run5 (fifth column), and run6 (sixth column). We show the dust disc evolution at times $t = 0\, \rm P_{orb}$ (first row), $t = 200\, \rm P_{orb}$ (second row), $t = 500\, \rm P_{orb}$ (third row), and $t = 1000\, \rm P_{orb}$ (fourth row). The dust disc is viewed in $y$-$z$ plane. The colour denotes the dust surface density. In every simulation, the emergence of dust rings is a result of the differential precession of gas and dust particles.}
\label{fig::splash}
\end{figure*}

We use the observed 99 Her binary parameters (given in the Introduction) to set up the binary star system. The 99 Her binary is modeled as sink particles with initial separation and accretion radius of  $a_{\rm b} = 16.5\, \rm au$ and $1.2a_{\rm b} = 20\, \rm au$, respectively, where $a_{\rm b}$ is the binary separation. Particles that cross the accretion radius deposit their mass and angular momentum onto the sink, which is regarded as a hard boundary. To shorten the computation time, the accretion radii of the stars are comparable to the binary separation, and particle orbits within the binary cavity are not resolved. A large accretion radius does not qualitatively change the evolution of the dust traffic jam \citepalias[see Section 3.5 in][]{Smallwood2024}. The simulations run for $1000\, \rm P_{orb}$, where $P_{\rm orb}$ is the binary orbital period of 99 Her. We extend run1 to $2000\, \rm P_{orb}$ to analyse longer-term disc structures.

The initial circumbinary disc comprises $500,000$ gas particles with equal masses and $50,000$ dust particles. These particles are distributed between the inner disc radius, approximately $r_{\rm in} = 40\, \rm au$ or $2.5a_{\rm b}$, and the outer disc radius, approximately $r_{\rm out} = 120\, \rm au$ or $7a_{\rm b}$. The gas surface density profile is initially a power-law distribution given by
 \begin{equation}
     \Sigma(r) = \Sigma_0 \bigg( \frac{r}{r_{\rm in}} \bigg)^{-p},
     \label{eq::sigma}
 \end{equation}
where $\Sigma_0 =  6\times10^{-3}\, \rm g/cm^2$ is the density normalization, $p$ is the power law index, and $r$ is the spherical radius. We use $p = +1$ and $+3/2$. The density normalization is determined based on the total mass of the disc, set to $M_{\rm d} = 0.001M_{\odot}$. For the specified total disc mass, we neglect the influence of disc self-gravity.  We use a locally isothermal equation-of-state given by
\begin{equation}
    c_{\rm s} = c_{\rm s,in} \bigg( \frac{r}{r_{\rm in}} \bigg)^{-q},
\end{equation}
where $c_{\rm s,in}$ is the sound speed evaluated at $r_{\rm in}$. The disc thickness is scaled with $r$ as
\begin{equation}
    H = \frac{c_{\rm s}}{\Omega} \propto r^{3/2-q}, 
 \end{equation}
where $\Omega = \sqrt{GM/r^3}$. We simulate two different values of $q$, $+1/2$ and $+3/4$.  We also simulate two different initial disc aspect ratio of $H/r = 0.05$ and $0.1$. In each case, we resolve the circumbinary disc with a shell-averaged smoothing length per scale height of $\sim \langle h \rangle/H \approx 0.34$ for $H/r = 0.05$ and $\langle h \rangle/H \approx 0.20$ for $H/r = 0.1$.

\begin{figure} 
\centering
\includegraphics[width=1\columnwidth]{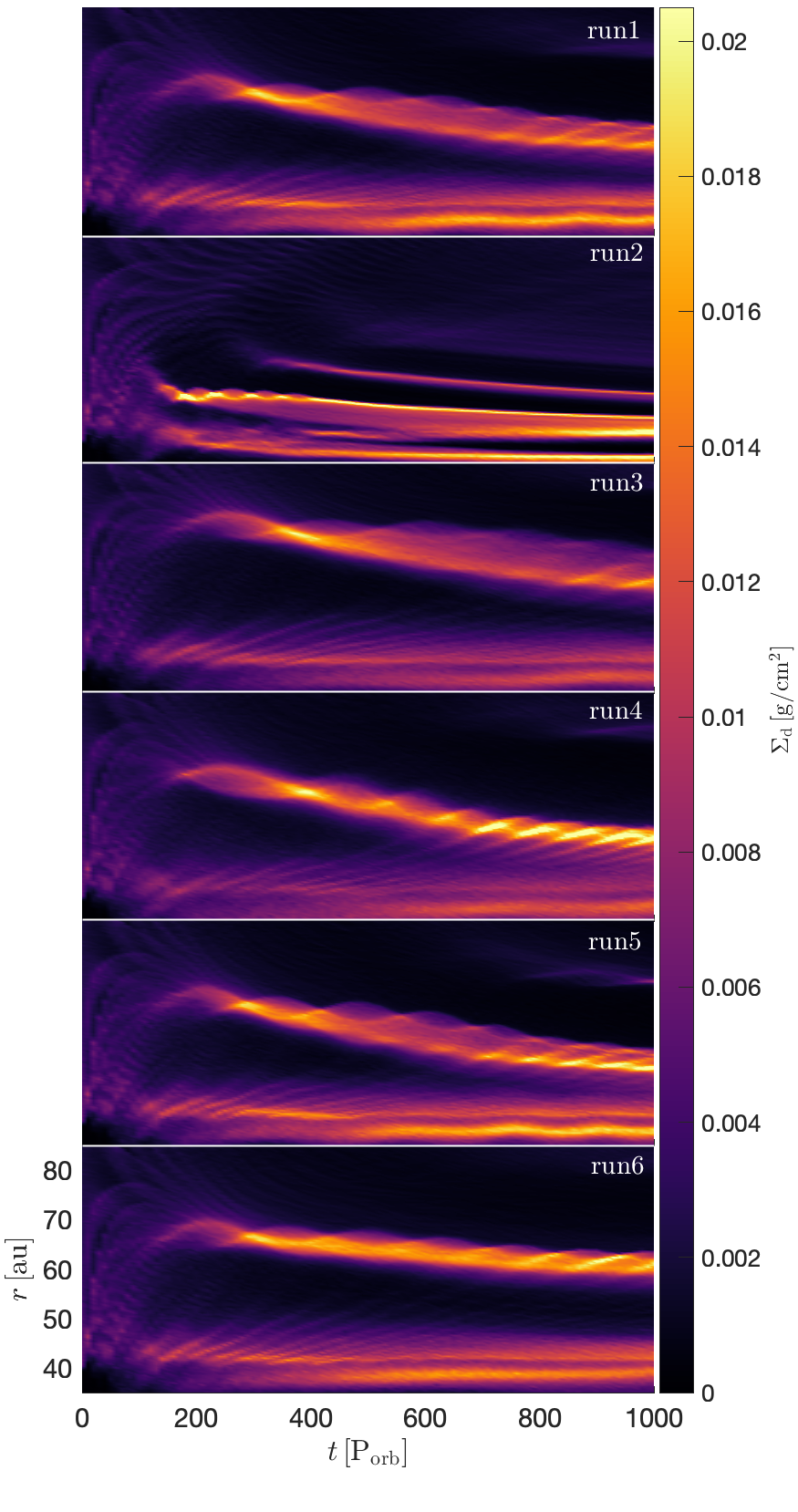}
\caption{The circumbinary disc dust surface density, $\Sigma_{\rm d}$, as a function disc radius, $r$, and time in binary orbital periods, $\rm P_{orb}$, for each hydrodynamical simulation.}
\label{fig::sigma}
\end{figure}

\begin{figure} 
\centering
\includegraphics[width=1\columnwidth]{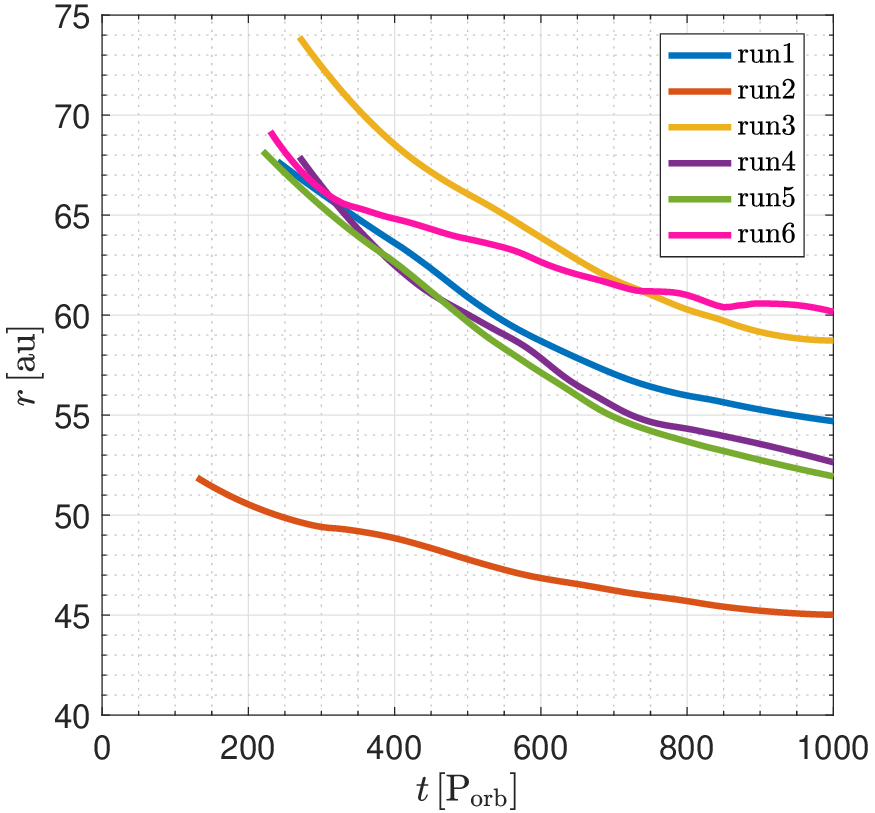}
\caption{A tracing of the peak dust surface density for the initial dust ring from Fig.~\ref{fig::sigma}  as a function of time for each simulation. }
\label{fig::peak}
\end{figure}

The \cite{Shakura1973} viscosity, $\alpha_{\rm SS}$, prescription is given by 
\begin{equation}
    \nu = \alpha_{\rm SS} c_{\rm s} H,
\end{equation}
where $\nu$ is the kinematic viscosity. To compute $\alpha_{\rm SS}$, we use the prescription in \cite{Lodato2010} given as
\begin{equation}
\alpha_{\rm SS} \approx \frac{\alpha_{\rm AV}}{10}\frac{\langle h \rangle}{H}.
\end{equation}
We simulate three values of the \cite{Shakura1973} viscosity parameter, $\alpha_{\rm SS} = 0.02$, $0.01$, and $0.005$, with an artificial viscosity $\alpha_{\rm AV} =  1$, $0.52$ and $0.26$, respectively. It is worth noting that a value of $\alpha_{\rm AV} =  0.1$ represents the lower limit below which the physical viscosity is not accurately resolved in SPH. In other words, viscosities smaller than this threshold lead to disc spreading independent of the value of $\alpha_{\rm AV}$ \cite[refer to][for more information]{Meru2012}. Additionally, the viscosity prescription includes a non-linear parameter, $\beta_{\rm AV}$, originally introduced to prevent particle penetration in high Mach number shocks \cite[e.g.,][]{Monaghan1989}, which is set to $\beta_{\rm AV} = 2.0$ \cite[e.g.,][]{Nealon2015}.


At the beginning of the simulation, the dust particles are initially distributed according to the same surface density profile as the gas. They have a dust-to-gas mass ratio of $0.01$. Each simulation adopts a uniform dust particle size. Specifically, we simulate an average Stokes number of approximately $65$, corresponding to particle sizes of $s = 3\, \rm cm$. Although the Stokes number varies with radius, the size of the dust particles remains constant throughout the simulation. The intrinsic density of the grains is assumed to be $3.00\, \rm g/cm^3$. Additionally, the initial aspect ratio of the dust particles matches that of the gas disc.

We systematically vary disc surface density, temperature, aspect ratio, and viscosity across all simulations while keeping other variables constant. This will allow us to discern the impact of each parameter on the formation and evolution of dust rings, facilitating a nuanced understanding of their interplay.


\begin{figure*} 
\centering
\includegraphics[width=1\columnwidth]{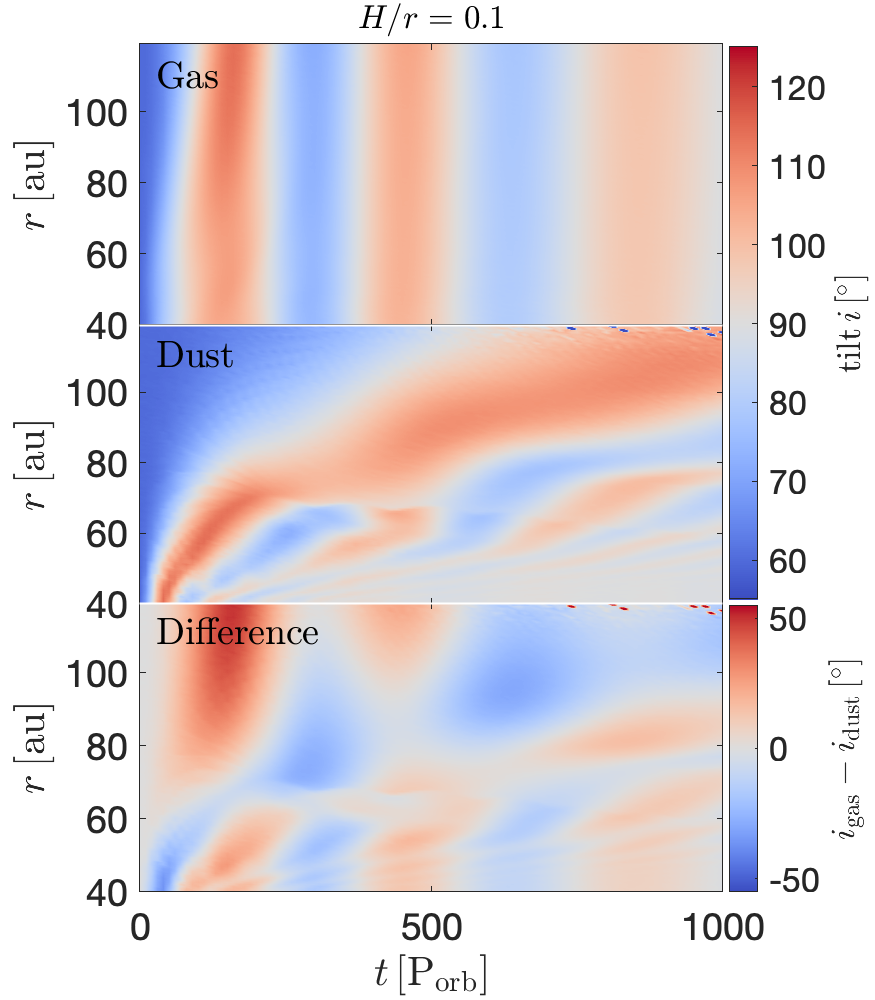}
\includegraphics[width=1\columnwidth]{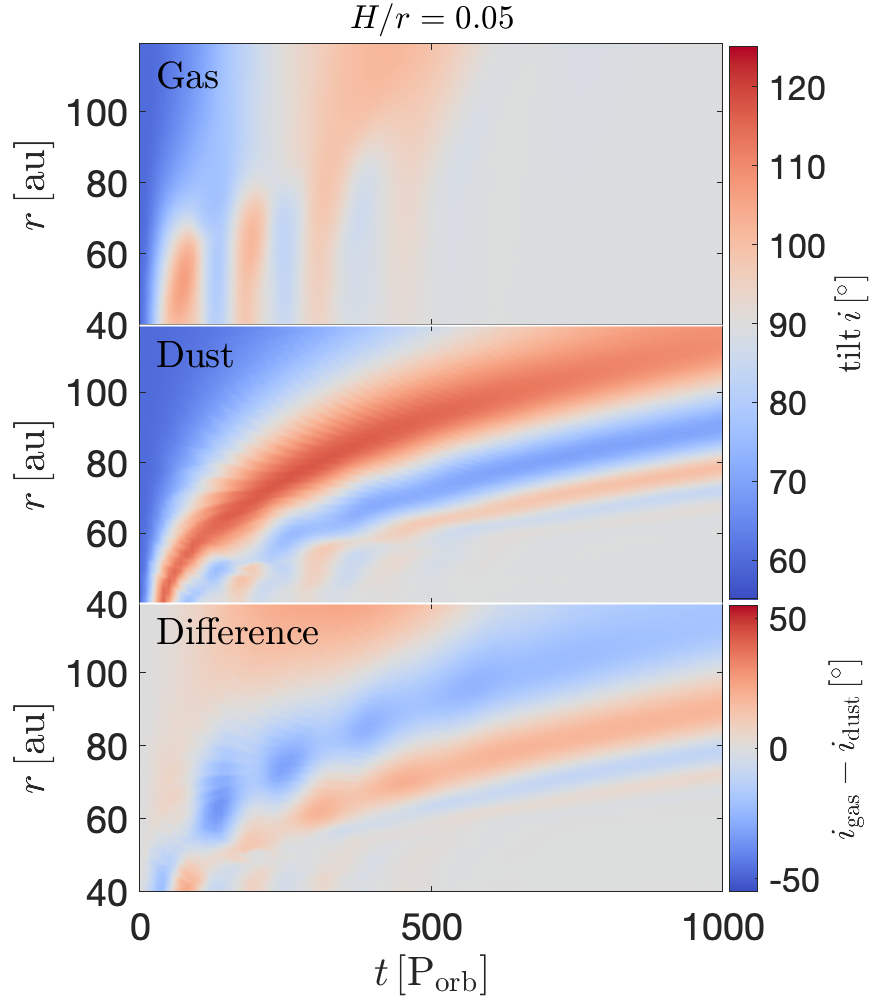}
\caption{The evolution of disc tilt for run1 ($H/r = 0.1$, left panel) and run2 ($H/r = 0.05$, right panel). The top two sub-panels present the tilt of the gas and dust, indicated by the colour bar, over time and radial distance. The discrepancy in tilt between gas and dust, represented as $i_{\rm gas} - i_{\rm dust}$, is shown in the lower sub-panel. Since the dust particles are drifting inward, there is lower resolution of dust particles in the outer parts of the disc. In each simulation,  initially (when $t < 200, \rm P_{orb}$), there is a notable difference in tilt between the gas and dust. By $t = 1000, \rm P_{orb}$, gas and dust are largely aligned in regions with $r \lesssim 60\, \rm au$, but misalignment persists for $r > 60\, \rm au$.}
\label{fig::tilt}
\end{figure*}

\section{Hydrodynamical results}
\label{sec::hydro_results}

Figure~\ref{fig::splash} illustrates the evolution of the dusty circumbinary disc surrounding 99 Her with varying disc aspect-ratio, initial surface density, temperature, and viscosity. The first row shows the initial dust disc structure viewed in the $y$--$z$ plane, where the colour represents the dust surface density. We then show the disc structure at times $t = 200\, \rm P_{orb}$, $500\, \rm P_{orb}$, and $1000\, \rm P_{orb}$, in the second, third, and last rows, respectively. Initially, the disc is inclined at an angle of $60^\circ$ but gradually aligns itself to a polar configuration. Even though we are just showing the dust, when the system reaches a time of $1000\, \rm P_{orb}$, the gaseous disc becomes nearly perpendicular to the orbital plane of the binary system. 

 In each simulation, a dust pileup occurs near the inner edge. This could be due to physical mechanisms where inward-drifting dust particles are halted at the gas inner edge or due to numerical resolution effects, where the better-resolved dust appears amplified as gas resolution diminishes \citep{Ayliffe2012}. Warp and alignment physics can be ruled out since this effect occurs in flat, planar discs as well \citep[Figure 4 in][]{Aly2021}. Resonances induced by the binary may also contribute, especially as the dust pileup intensifies with higher Stokes numbers \cite[e.g.,][]{Smallwood2024}. In Appendix~\ref{app::pressure}, a pressure bump is present near the inner disc edge at early times, whereas at the location of each dust traffic jam there are no pressure bumps present. Thus, this phenomenon cannot be attributed to a 'traffic jam' effect, but its exact origins remain unclear and require further investigation.

The first column in Fig~\ref{fig::splash} shows our control simulation (run1), where the initial dust traffic jam (the outer ring) occurs at $t \sim 200\, \rm P_{orb}$ due to the differential precession between the gas and dust \citep{Aly2020,Aly2021,Smallwood2024}.  The inner dust rings near the inner disc edge are not formed by the differential precession between the gas and dust (see Appendix~\ref{app::pressure}). As the disc evolves, the dust traffic jams evolve into a coherent dust ring. Throughout the simulation, the dust ring gradually drifts inward.  The two-armed spiral structure in the dust is related to the spirals launched in the gas by the binary potential, and also may be affected by the formation of the dust traffic jams (see Appendix~\ref{app::spiral}). When we decrease the $H/r$ (second column, run2), the dust ring forms nearer to the binary.  We note that for this value of $H/r$, the disc breaks during polar alignment, which may impact the dynamics of the dust traffic jams \cite[e.g.,][]{Aly2021}. We show further evolution of this case in Appendix~\ref{app::disc_mor}. Decreasing $p$ (third column, run 3) causes the initial dust ring to form at a further radial distance. On the contrary, decreasing $q$ (fourth column, run 4)  shows the simulation structure to the control simulation. Likewise, when we vary the $\alpha$ parameter, the dust ring structure is consistent with the control simulation. The ring position trends align with the analytical findings presented in \cite{Longarini2021}, as illustrated in their Figure 2.



\begin{figure*} 
\centering
\includegraphics[width=1\columnwidth]{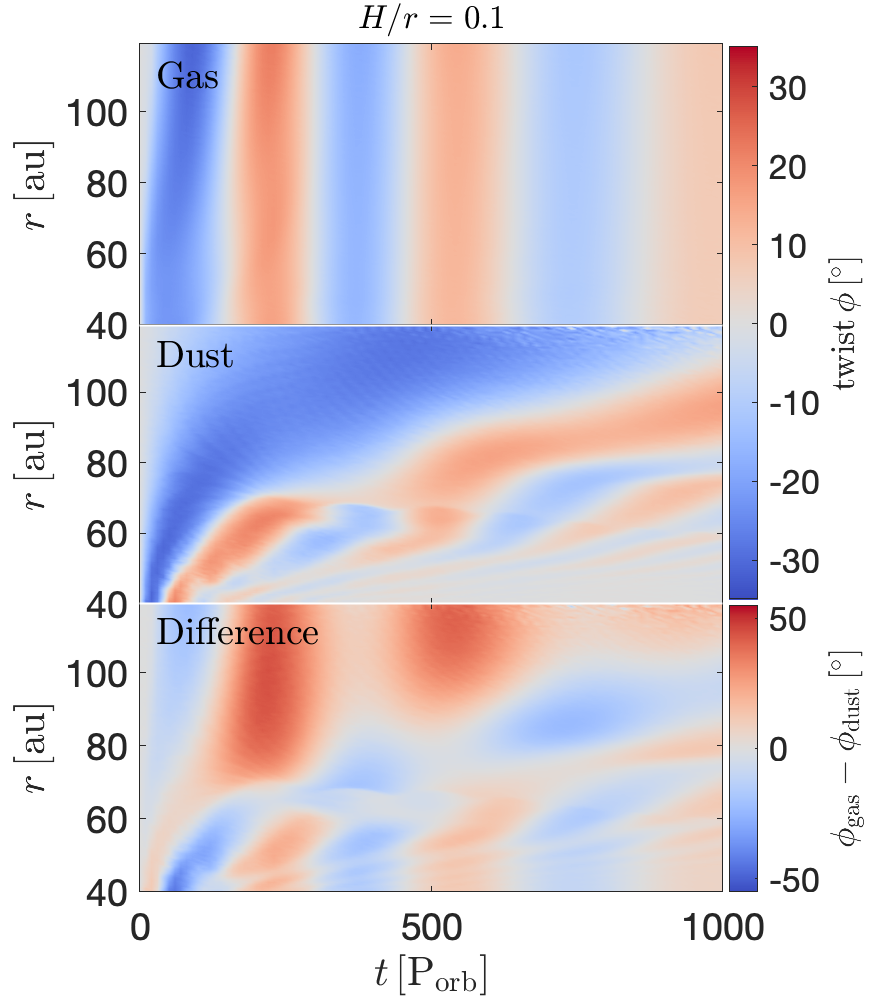}
\includegraphics[width=1\columnwidth]{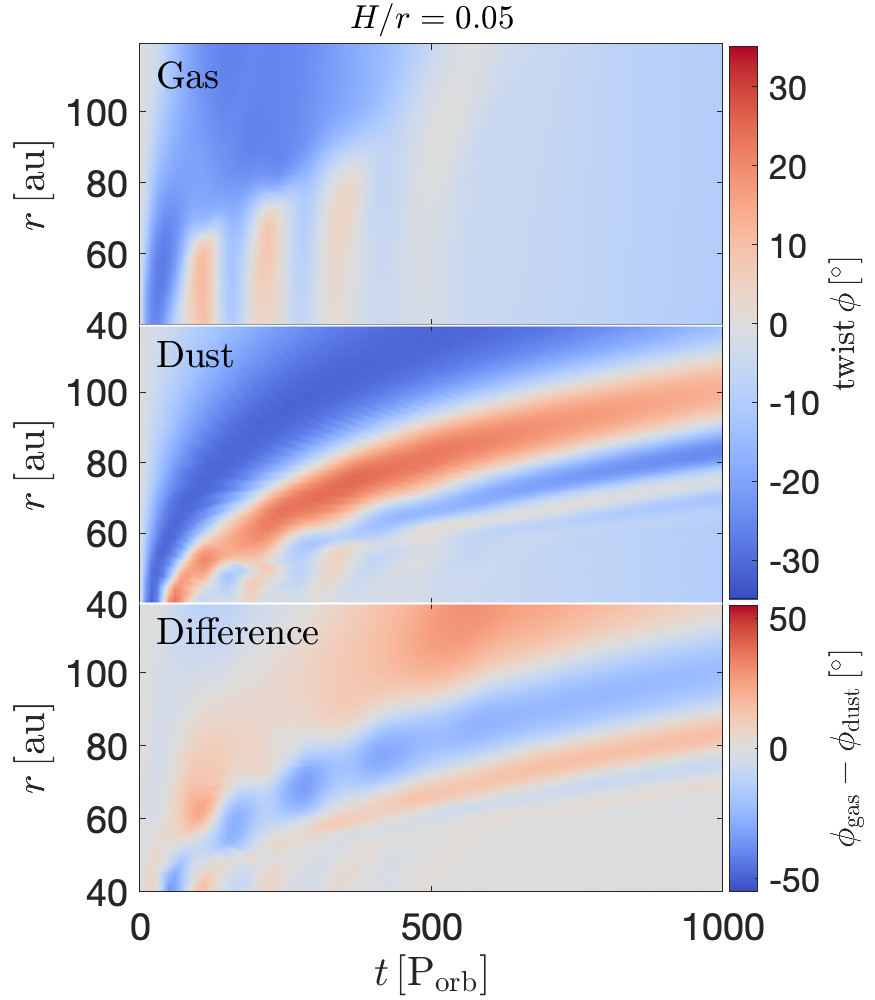}
\caption{The evolution of the disc's longitude of the ascending node for run1 ($H/r = 0.1$, left panel) and run2 ($H/r = 0.05$, right panel). The top two sub-panels present the longitude of the ascending node, or twist, of the gas and dust, indicated by the colour bar, over time and radial distance. The discrepancy in the longitude of the ascending node between gas and dust, represented as $\phi_{\rm gas} - \phi_{\rm dust}$, is shown in the lower sub-panel. In each simulation,  initially (when $t < 200, \rm P_{orb}$), there is a notable difference in $\phi$ between the gas and dust. By $t = 1000, \rm P_{orb}$, gas and dust are largely aligned in regions with $r \lesssim 65, \rm au$, but misalignment persists for $r > 65, \rm au$. Since the dust particles are drifting inward, there is lower resolution of dust particles in the outer parts of the disc. The misalignment of gas and dust will undergo differential precession, which continues to generate dust traffic jams. }
\label{fig::twist}
\end{figure*}

To further analyse the structure and evolution of the dust ring in each simulation, we show the temporal evolution of the dust surface density as a function of disc radius in Fig.~\ref{fig::sigma}. In every model, the initial formation of the dust ring occurs within a period of approximately $150-300$ orbital periods ($\rm P_{orb}$) and subsequently migrates inward. The width of the dust ring remains consistent across runs 1, 3, 4, and 6. However, a narrower dust ring materializes in the case of decreased $H/r$ (run 2), and  decreased $\alpha$ value (run 5). Figure~\ref{fig::peak} illustrates the tracking of the highest dust surface density peak from Fig.~\ref{fig::sigma}, depicting its temporal behavior for each simulation. The dust rings in runs 1, 4, 5, and 6 all exist around the same time and at identical radial distances within the disc. For the scenario with $H/r = 0.05$, the initial dust ring takes shape earlier at approximately $150\, \rm P_{orb}$ and within a narrower radial span of around $53$ $\rm au$. Conversely, when the power law index of the surface density, denoted as $p$, is reduced to $p = 1$, the dust ring forms at a larger radius, approximately $75$ $\rm au$. Increasing $\alpha$ causes the dust ring to migrate inward more slowly.

Figure~\ref{fig::tilt} displays the disc tilt evolution as a function of time. The tilt profiles for the simulations with $H/r=0.1$ show similar behaviour. Therefore, we only show the tilt evolution for varying $H/r$, run 1 ($H/r = 0.1$, left panel) and run 2 ($H/r= 0.05$, right panel).  The upper two sub-panels show the tilt of both the gas and dust (represented by the colour bar) over time and radial distance. For the simulation with $H/r = 0.1$, the gas disc aligns to a polar configuration, $i_{\rm gas} \sim 90^\circ$, as a rigid body. When $H/r = 0.05$, the inner parts of the disc align polar on a faster timescale than the outer regions of the disc.  The disc alignment timescale increases with increasing $H/r$. Therefore, a thinner disc will align polar on a faster timescale \cite[e.g.,][]{Lubow2018}. During disc alignment, the gas experiences oscillations in tilt prompted by the influence of 99 Her. The dust also experiences oscillations in tilt during the alignment process. Dust particles within the inner disc regions (below $60\, \rm au$) gradually align toward an almost polar configuration. On the other hand, dust in the outer disc regions (beyond $60\, \rm au$) remains out of alignment with the gas throughout the simulation duration. The disparity in tilt between the gas and dust, expressed as $i_{\rm gas} - i_{\rm dust}$, is presented in the lower sub-panel. In the initial stages (when $t < 200, \rm P_{orb}$), the tilt difference between the gas and dust is substantial. By the time $t = 1000, \rm P_{orb}$, the gas and dust are largely aligned in regions where $r \lesssim 60\, \rm au$, but continue to be misaligned for $r > 60\, \rm au$.

Similar to Fig.~\ref{fig::tilt}, Fig.~\ref{fig::twist} depicts the progression of the disc's longitude of the ascending node as a function of time. The twist profiles for the simulations with $H/r = 0.1$ show similar behaviour. Therefore, we only show the twist evolution for varying $H/r$, run 1 ($H/r = 0.1$, left panel) and run 2 ($H/r= 0.05$, right panel). In the simulation with $H/r = 0.1$, the gas disc undergoes precession around the eccentricity vector, resembling precession of a rigid body. When considering $H/r = 0.05$, the disc's inner portions exhibit quicker precession than the outer regions. Analogous to the tilt evolution, the precession behavior of the dust differs from that of the gas. Unlike gas, dust does not process uniformly, which is vital in generating dust traffic jams within the disc. The difference in the longitude of the ascending node between the gas and dust, represented as $\phi_{\rm gas} - \phi_{\rm dust}$, is displayed in the lower sub-panel. When $t < 200, \rm P_{orb}$, the twist difference between the gas and dust is substantial. By the time $t = 1000, \rm P_{orb}$, the gas and dust are largely aligned in regions where $r \lesssim 60\, \rm au$, but continue to be misaligned for $r > 60\, \rm au$, similar to the disc tilt profile.

The difference between the tilt (lower panel of Fig,~\ref{fig::tilt}) and twist (lower panel of Fig.~\ref{fig::twist}) both contribute to the formation of the dust traffic jams. When the precession of a circumbinary disc has a component around the eccentricity vector, then the precession is reflected through the tilt and twist. For a circular binary precessing around the binary angular momentum vector, the precession is reflected merely by the twist, and not the tilt. For a circumbinary disc undergoing polar alignment, the precession is reflected through tilt oscillations more so than the twist \citep{Aly2021}. For both the tilt and twist differences in Figs.~\ref{fig::tilt} and~\ref{fig::twist}, as time progresses, the difference diminishes as gas and dust align. Nonetheless, even at $t = 1000, \rm P_{orb}$, regions within the disc still experience significant differences is the gas and dust tilt and twist, contributing to the ongoing generation of dust traffic jams.


\section{Streaming instability growth timescales}
\label{sec::Streaming_instability}

\subsection{Model setup}
Dust within protoplanetary discs, including circumbinary discs, can grow to millimeters to centimeter sizes through processes such as agglomeration, but further growth is impeded by bouncing or fragmentation \citep{Blum2018,Dominik2024}. A swarm of collectively self-gravitating particles can undergo direct gravitational collapse into planetesimals to overcome the bouncing and fragmentation barrier. However, for direct collapse to occur, the dust density must be large relative to the gas \citep{Shi2013}. Large dust-to-gas ratios can be attained through dust settling, radial drift, particle trapping, or other dust-gas instabilities \citep{Chiang2010,Johansen2014,Lesur22}, including the streaming instability \cite[SI;][]{Youdin2005}.



Numerical simulations demonstrate that the SI effectively initiates the direct gravitational collapse of dust clumps \citep{Johansen2009,Carrera2015,Simon2016,Simon2017,Schafer2017,Nesvorny2019,Li2021},  given sufficient particle size and the local dust-to-gas ratio mass density is of order unity or larger. Here, we make a first assessment of whether or not planetesimals can form via the SI in our simulated systems. While current SPH simulations cannot resolve the SI,  we can estimate SI growth rates given the midplane dust-to-gas ratios and particle sizes
in our simulations. To this end, we numerically solve the linearized, two-fluid equations in the shearing box approximation \citep[e.g.][]{Chen2020}. For simplicity, we neglect viscosity here (i.e., conduct inviscid calculations). This is the standard scenario considered in previous analytic SI calculations \citep[][]{Youdin2005,Kowalik2013}.  \cite{Chen2020} found that inducing the SI in nonlaminar protoplanetary discs, particularly for small particles, poses significant challenges.


\begin{figure} 
\centering
\includegraphics[width=\columnwidth]{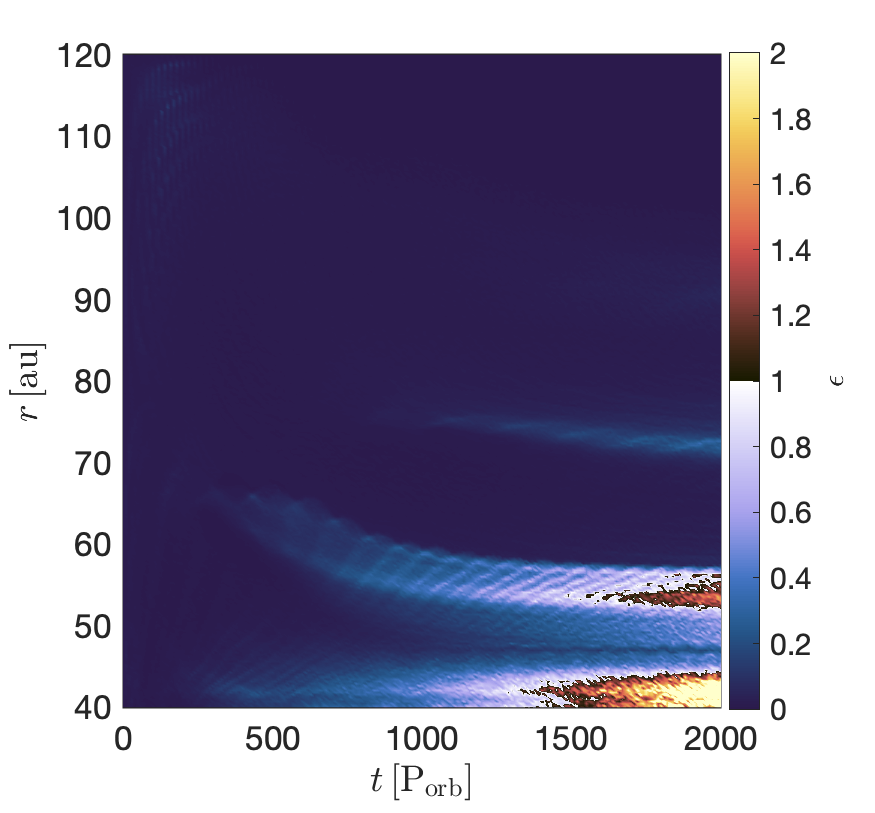}
\caption{The azimuthally-averaged circumbinary disc midplane dust-to-gas ratio, $\epsilon$, for our control simulation (run1) as a function of disc radius, $r$, and time in binary orbital periods, $\rm P_{orb}$. As the dust ring evolves in time, $\epsilon$ becomes greater than unity which can trigger the streaming instability.}
\label{fig::mid_ctrl}
\end{figure}

\begin{figure} 
\centering
\includegraphics[width=\columnwidth]{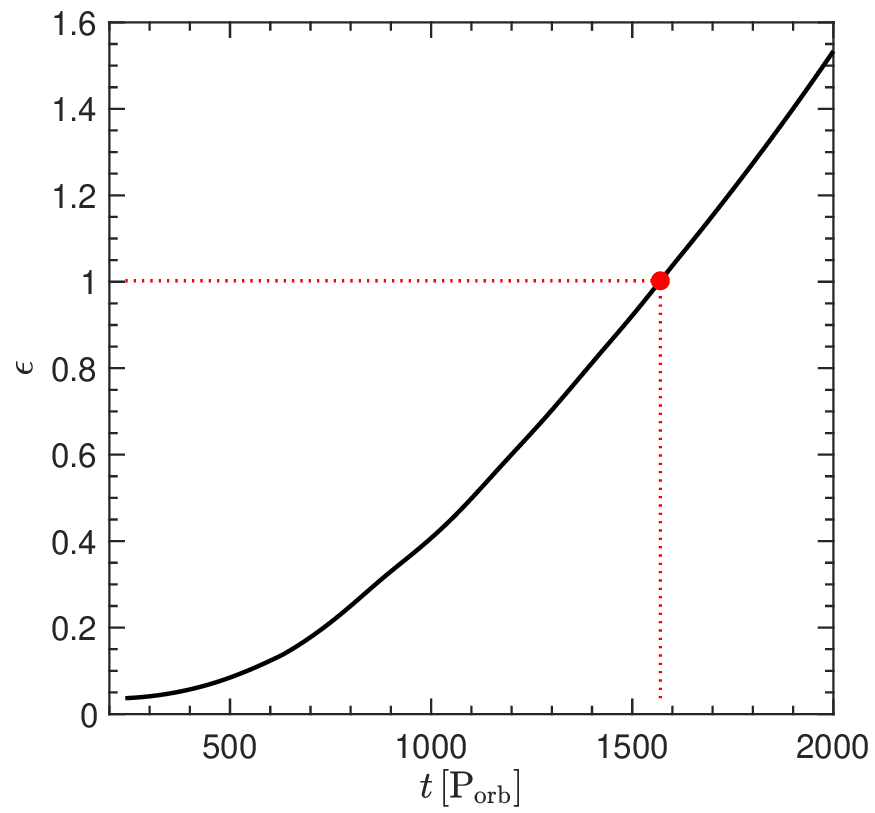}
\caption{We trace the midplane dust-to-gas ratio, $\epsilon$, along the initial dust ring as a function of time in binary orbital periods, $\rm P_{orb}$. The red dot highlights the moment when  $\epsilon \simeq 1$, occurring at $t \approx 1600, \rm P_{orb}$. As time progresses, $\epsilon$ increases, approaching $\epsilon \approx 1.6$ by the conclusion of the simulation.}
\label{fig::mid_time}
\end{figure}


Before continuing with the analytical calculations, we further analyse the SPH simulations to retrieve the dust-to-gas ratio at the midplane, $\epsilon = \rho_{\rm d}/\rho_{\rm g}$. We use our standard simulation, extended to $2000\, \rm P_{orb}$. The disc is still sufficiently resolved at this time, with $\langle h \rangle/H \approx 0.35$. 
From \citetalias{Smallwood2024}, the authors showed that the dust-to-gas ratio in the midplane within the dust ring could exceed unity, which may be a favorable place for the streaming instability to operate. Therefore, we show the analysis of the dust-to-gas ratio.
Figure~\ref{fig::mid_ctrl} shows the azimuthally-averaged disc midplane dust-to-gas ratio as a function of disc radius and time. As the dust ring evolves, $\epsilon$ increases in time. Beginning at approximately $t \sim 1600\, \rm P_{orb}$, $\epsilon$ exceeds unity in the rings. 
Figure~\ref{fig::mid_time} shows the midplane dust-to-gas ratio, $\epsilon$, along the initial dust ring as a function of time in binary orbital periods. Throughout the simulation $\epsilon$ consistently rises within the dust ring. At approximately $t \sim 1600, \rm P_{orb}$, $\epsilon$ achieves unity. By the conclusion of the simulation, $\epsilon$ approaches 1.6. In numerical simulations, surpassing a unity ratio in the midplane can trigger the streaming instability, leading to substantial clumping \cite[e.g.,][]{Johansen2009,Schafer2017,Nesvorny2019,Li2021}.

\begin{figure*} 
\centering
\includegraphics[width=\columnwidth]{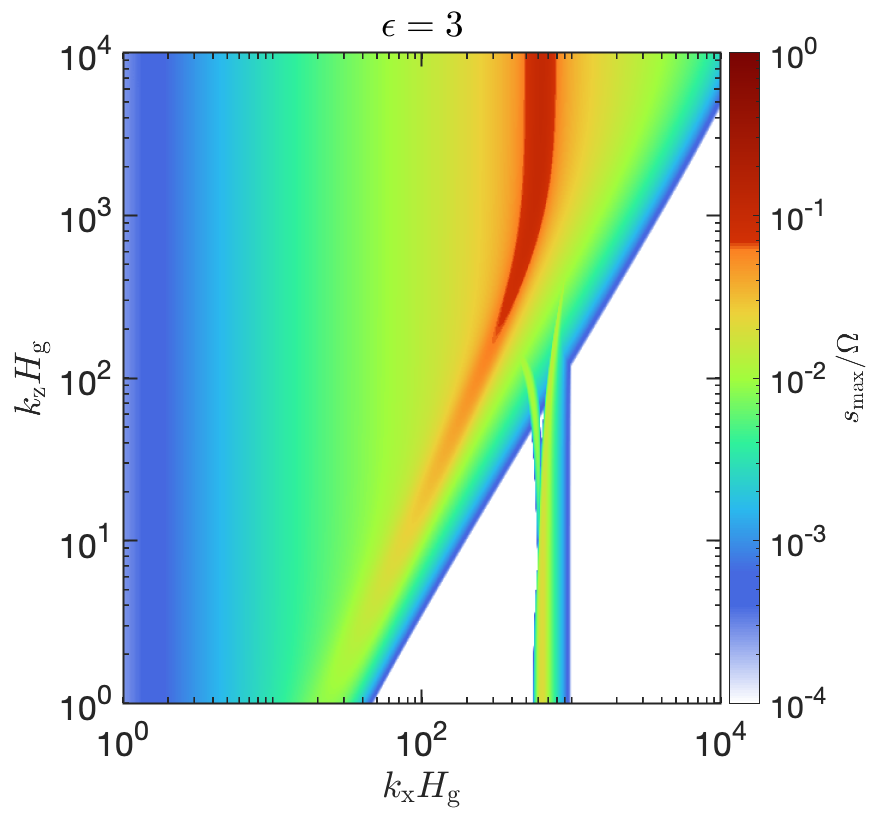}
\includegraphics[width=\columnwidth]{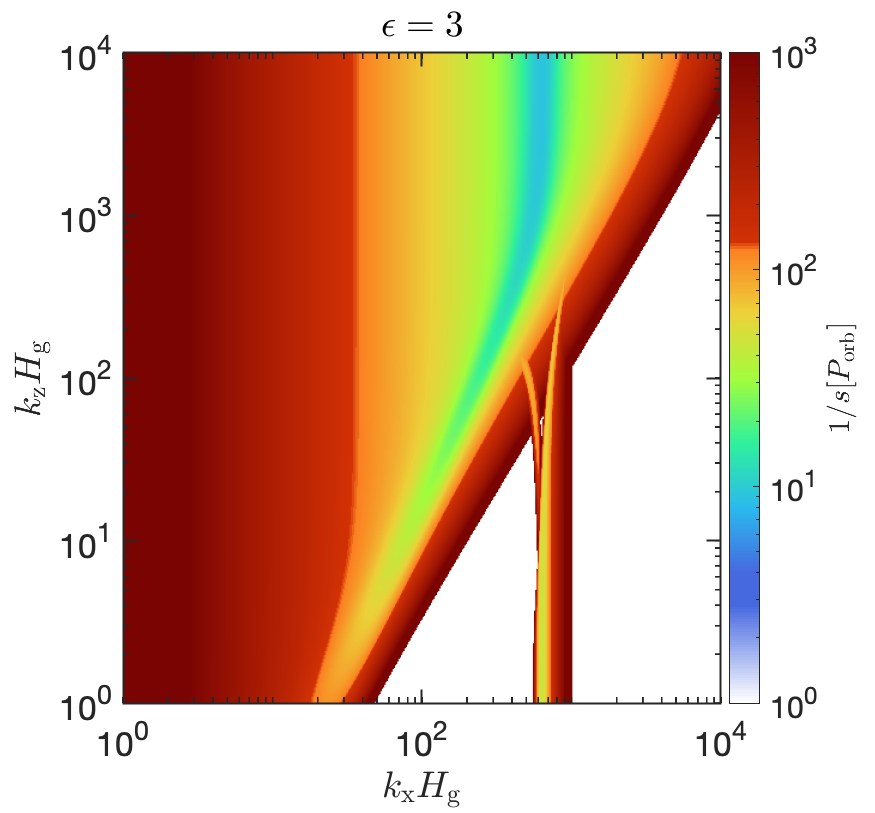}
\includegraphics[width=\columnwidth]{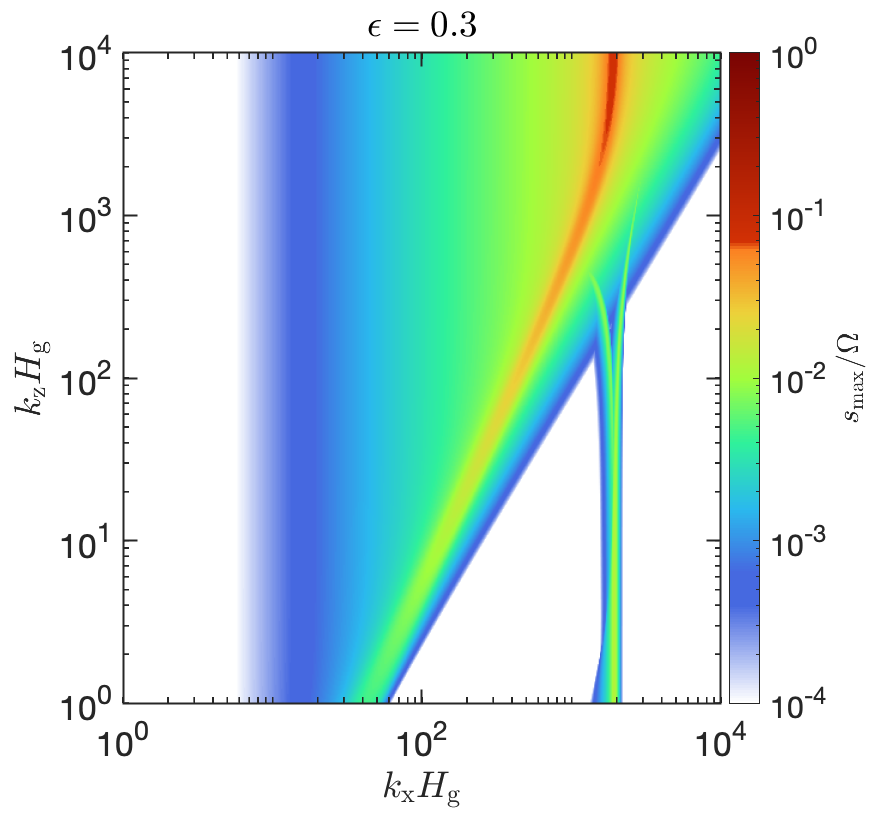}
\includegraphics[width=\columnwidth]{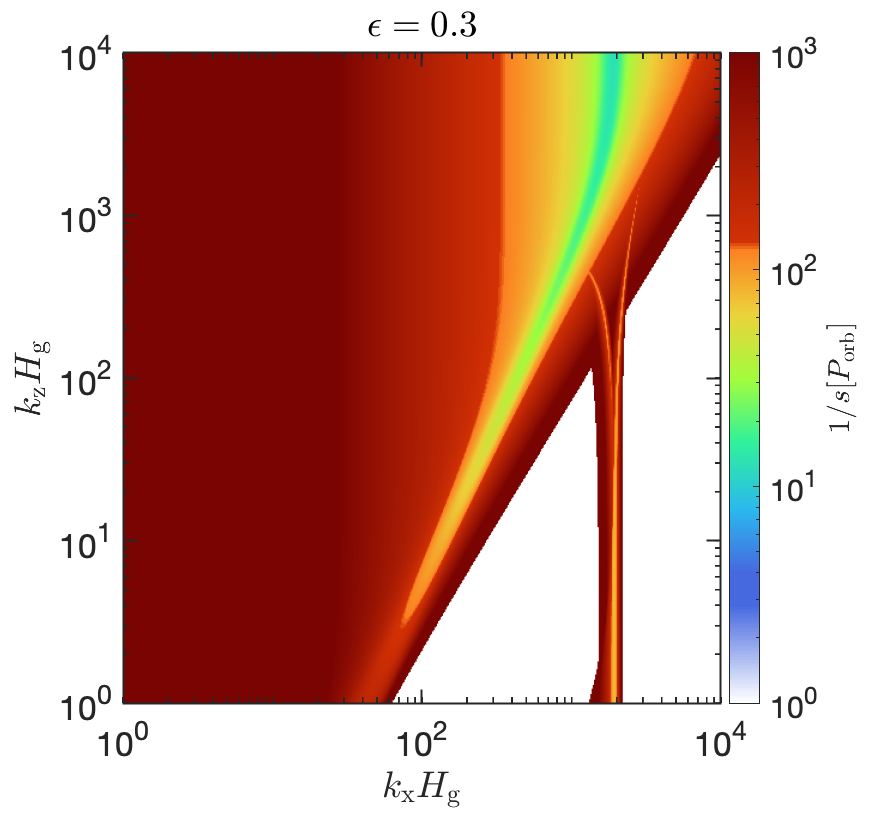}
\caption{The  unstable modes in dusty discs showing the streaming instability growth rates as a function of wavenumbers with $\epsilon = 3$ (top-left) and $\epsilon = 0.3$ (bottom left),  at $t = 1600\, \rm P_{orb}$ which corresponds to $\rm St \sim 250$. We show the streaming instability growth timescale in units of 99 Her orbital period ($P_{\rm orb}$) as a function of wavenumbers for a dusty disc with $\epsilon = 3$ (top-right) and  $\epsilon = 0.3$ (bottom-right), both at $r = 55\, \rm au$  (the radial location of the dust ring from the simulation)}.
\label{fig::kx_kz_eps3}
\end{figure*}

In this model, the dust is treated as a second, pressureless fluid interacting with gas via drag forces characterized by a single stopping time or Stokes number $\st$. The gas and dust velocities in the shearing box, relative to the background Keplerian shear flow, are denoted by $\bm{v}_d$ and $\bm{v}_g$, respectively. The SI is powered by the relative motion between the two species in the equilibrium disc, $\Delta v \equiv \bm{v}_d - \bm{v}_g$. In the classic SI, the radial ($x$) and azimuthal ($y$) components are given by:
\begin{align}
    \Delta v_{\rm x} & = - \frac{2\st(1+\epsilon)}{\Delta^2}\eta R\Omega_{\rm k},\\
    \Delta v_{\rm y} & = \frac{\st^2}{\Delta^2}\eta R\Omega_{\rm k}, 
\end{align}
while $\Delta v_{\rm z} = v_{\rm dz} = v_{\rm gz}=0$, where 
\begin{equation}
    \Delta^2 = \rm St^2 + (1+\epsilon)^2
\end{equation}
and
\begin{equation}
    \eta \equiv - \frac{1}{2 R \rho_\mathrm{g}\Omega^2} \pdv{P}{R}
\end{equation}
is a dimensionless measure of the global radial pressure gradient \citep{Youdin2005}.  The above equations follow from equations 13 - 15 of \cite{Youdin2005}. Typically, $\eta$ is $O(h^2)$, and in the linear problem, only the parameter $\eta/h$ is relevant, where $h$ is the aspect ratio. We set $\eta/h =0.05$ (a standard value) in the examples below.  We also analyse the system at $t = 1600\, \rm P_{orb}$ and $r=55\, \rm au$ (the location of the dust ring), which corresponds to $\epsilon \sim 1$ and $\rm St\sim 250$. Before proceeding, it is essential to note that by adopting the above equilibrium, we have assumed that the radial pressure gradient dominates the dust-gas drift. However, in a distorted disc such as the one undergoing polar alignment, there are other sources of dust-gas drift, e.g., differential precession. \cite{Aly2021} found that the differential precession may be higher than the dust-gas drift, unless close to the co-precession radius. We leave this to future work.


We perturb the two-fluid system with axisymmetric Eulerian perturbations that depend on the radial and vertical wavenumbers, $k_{\rm x,z}$ (taken to be positive without loss of generality), and the complex frequency $\sigma$ with growth rate $s  \equiv  \rm Re(\sigma)$. We solve the linearized equations and find the dimensionless SI growth rate $S \equiv s/\Omega$ (using equations 26 - 29 of \cite{Youdin2005}), where $\Omega = \sqrt{G (M_1 + M_2)/R^3}$ is the Keplerian angular frequency. We set $\epsilon=3$ and $\epsilon=0.3$  as representative values in the dust rings, but note that the results below are not sensitive to values of $\epsilon$ of order unity. The value of $\epsilon$ from Fig.~\ref{fig::mid_time} is in between our selected values of $\epsilon$. Note, $\epsilon\sim0.3$ is the peak dust-to-gas ratio when considering a more realistic sink accretion radius (see Fig.~16 in \citetalias{Smallwood2024}). 

\subsection{Application to 99 Her}
We apply the SI model to the hydro simulations with application to 99 Her. In the top-left panel in Fig.~\ref{fig::kx_kz_eps3}, we plot dimensionless growth rates as a function of wavenumbers for $\epsilon=3$. We find SI modes with growth rates up to $\log(s_{\rm max}/\Omega) \sim -0.5$ for $k_{\rm z}H_g  \gtrsim 1$. We note that $ \frac{\Delta v_{\rm y}}{\Delta v_{\rm x}} \propto \frac{\st}{(1+\epsilon)}$, thus in the limit of $\st\gg 1$ the dust-gas drift is dominated by azimuthal drift rather than radial drift. This suggests that instability is powered by $\Delta v_y$ for large $\st$. We verified this by artificially setting $\Delta v_x = 0$ , and obtaining similar growth rates as when we include $\Delta v_x$.  Azimuthal drift also powers the "vertical" branch of unstable modes with $k_{\rm z}H_g \lesssim 10$ and $k_xH_g\sim 100$. In fact, this branch persists even for $k_{\rm z} \equiv 0$ and has previously been identified in accreting disc models as the "azimuthal drift" SI \citep[see][]{Lin2022}. The bottom-left panel in Fig.~\ref{fig::kx_kz_eps3} shows the dimensionless growth rates as a function of wavenumbers for $\epsilon=0.3$. For a dust-to-gas ratio less than unity, we still find SI modes with growth rates up to $\log(s_{\rm max}/\Omega) \sim -0.5$ for $k_{\rm z}H_g  \gtrsim 1$ but with a smaller range in $k_{\rm x}$ compared to $\epsilon=3$.

In the top-right panel in Fig.~\ref{fig::kx_kz_eps3}, we plot SI growth timescale as a function of wavenumbers for $\epsilon=3$.  Since our hydrodynamical simulations show the dusty ring drifts inwards, we probe the growth timescales for $r = 55 \rm au$. We use the Keplerian angular frequency around the binary 99 Her to convert the dimensionless growth timescale into units of binary orbital period, $P_{\rm orb}$. The growth timescale is $\lesssim 10\, \rm P_{\rm orb}$. Therefore, the timescale for the growth induced by the SI, is less than the tilt oscillation timescale during the alignment process. For the 99 Her system, the circumbinary gas disc aligns on a timescale $\sim 1000\, \rm P_{\rm orb}$. The bottom-right panel in Fig.~\ref{fig::kx_kz_eps3} shows the SI growth timescale as a function of wavenumbers for $\epsilon=0.3$.  For a dust-to-gas ratio less than unity, we still find a SI growth timescale of $\lesssim 10\, \rm P_{\rm orb}$ but with a smaller range in $k_{\rm x}$ compared to $\epsilon=3$. From our simulations, the dust ring is still present once the gas disc is polar, suggesting growth induced by the SI may aid in the formation of polar planets.  However, we note that if the growth proceeds too rapidly, the Stokes number will increase before the grains reach polar alignment. This could make it more difficult for the solids to follow the gas to a polar state.



\section{Summary}
\label{sec:Summary}
We investigated the formation of dust traffic jams in polar-aligning circumbinary discs with application to the 99 Herculis binary system. We extended the work of \citetalias{Smallwood2024}, modeling 3D smoothed particle hydrodynamical simulations of an initially highly inclined circumbinary disc with gas and dust components. Unlike previous works on this topic, we focused on how the disc properties affected the formation and evolution of the dust traffic jams. These properties included the disc aspect ratio, surface density power law index, temperature power law index, and viscosity. For each disc property, dust traffic jams within the disc are produced by the differential precession between gas and dust components as the disc aligns to a polar state. These polar dust rings are long-lived and robust to the different disc parameters chosen here. The inward migration of these dust rings, coupled with midplane dust-to-gas ratios exceeding unity, highlights the potential role of the streaming instability in fostering conditions conducive to the formation of polar planets.

We followed up our SPH simulations with analytical calculations of streaming instability growth rates and growth timescales. We find unstable modes in a dusty disc for $\rm St > 1$. We then used the Keplerian angular frequency around the binary 99 Her to convert the dimensionless growth timescale into units of the binary orbital period. We found the streaming instability growth timescale is less than the tilt oscillation timescale during the alignment process. Therefore, the dust ring will survive once the gas disc aligns polar, suggesting that the streaming instability will aid in forming polar planets.

The formation of dust rings in polar-aligning circumbinary discs has implications for the formation of polar circumbinary planets. 
This work showed that a sufficiently misaligned gaseous and dusty circumbinary disc around an eccentric binary will undergo polar alignment within the gas disc lifetime -- forming a stable dusty polar disc. We found dust rings are produced during the alignment process, and are stable even when the circumbinary disc is in a polar configuration. From Section~\ref{sec::Streaming_instability}, we found that these dust rings may a favourable environment for the streaming instability to operate, which can form planetesimals. Eventually, these planetesimals will form planetary cores through the core accretion model of planet formation \citep{Safronov1969,Goldreich1973}. This implies that planets forming in a polar circumbinary disc would have polar orbits around the binary. There may be evidence for the first polar planet around the post-AGB binary AC Her \citep{Martin2023}. However, there are no confirmed detections of polar circumbinary planets. 

If polar circumbinary planets can form, which, given our models, is likely, they would be harder to detect than nearly coplanar circumbinary planets found by recurrent transits using Kepler. Polar planets may be detectable as non-recurring transits of the binary or eclipse timing variations of the binary \cite[e.g.,][]{ZhangFabrycky2019}. The most compelling binary system to host polar circumbinary planets is 99 Her. The best-fitting model to reproduce the 99 Her debris disc observations is a thin polar debris ring \citep[e.g.,][]{Kennedy2012}. A polar debris disc may be shaped into a thin ring by the presence of polar planets through a process of shepherding planets \citep{Rodigas2014}. 


\section*{Acknowledgements}
 We thank the referee for helpful comments that improved the quality of this manuscript. JLS acknowledges funding from the ASIAA Distinguished Postdoctoral Fellowship. MKL
is supported by the National Science and Technology Council
(grants 111-2112-M-001-062-, 112-2112-M-001-064-, 111-2124-M-002-013-, 112-2124-M-002 -003-) and an Academia Sinica Career Development Award (AS-CDA110-M06). RN acknowledges funding from UKRI/EPSRC through a Stephen Hawking Fellowship (EP/T017287/1). HA acknowledges funding from the European Research Council (ERC) under the European Union’s Horizon 2020 research and innovation programme (grant agreement No 101054502). CL aknowledges fundings from the European Union’s Horizon 2020 research and innovation programme under the Marie Skłodowska-Curie grant agreement No 823823 (RISE DUSTBUSTERS project).

\section*{Data Availability}
The data supporting the plots within this article are available on reasonable request to the corresponding author. A public version of the {\sc phantom} and {\sc splash} codes are available at \url{https://github.com/danieljprice/phantom} and \url{http://users.monash.edu.au/~dprice/splash/download.html}, respectively.



\bibliographystyle{mnras}
\bibliography{ref.bib} 




  \appendix

  \section{Differentiating between dust pileups versus traffic jams}
  \label{app::pressure}

 Figure~\ref{app::pressure} shows the gas and dust surface density and the pressure gradient as a function of radius. At a time $t = 170\, \rm P_{orb}$ (left panel), a pressure bump near the inner disc edge is present which causes the dust to pileup at this location. The dust accumulation at $r\sim 65\, \rm au$ is produced by a traffic jam from the differential precession between the gas and dust, as there is no pressure bump at this location. At a time $t = 1600\, \rm P_{orb}$ (right panel), the two outer dust accumulations are again driven by differential precession since no pressure bumps are present.

 \begin{figure*} 
\centering
\includegraphics[width=1\columnwidth]{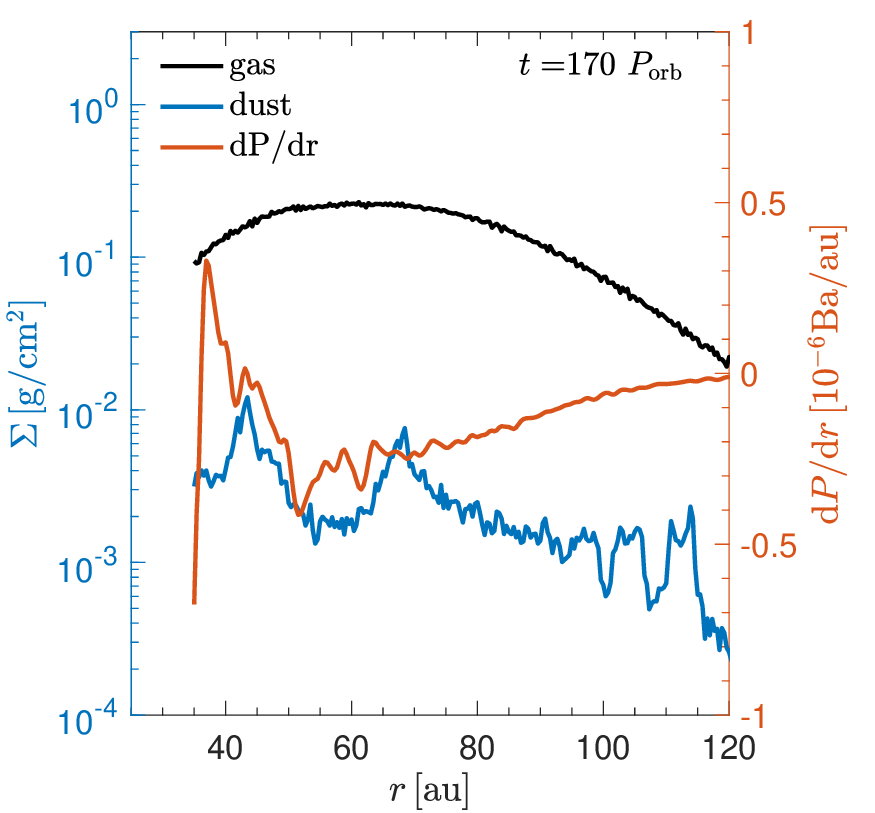}
\includegraphics[width=1\columnwidth]{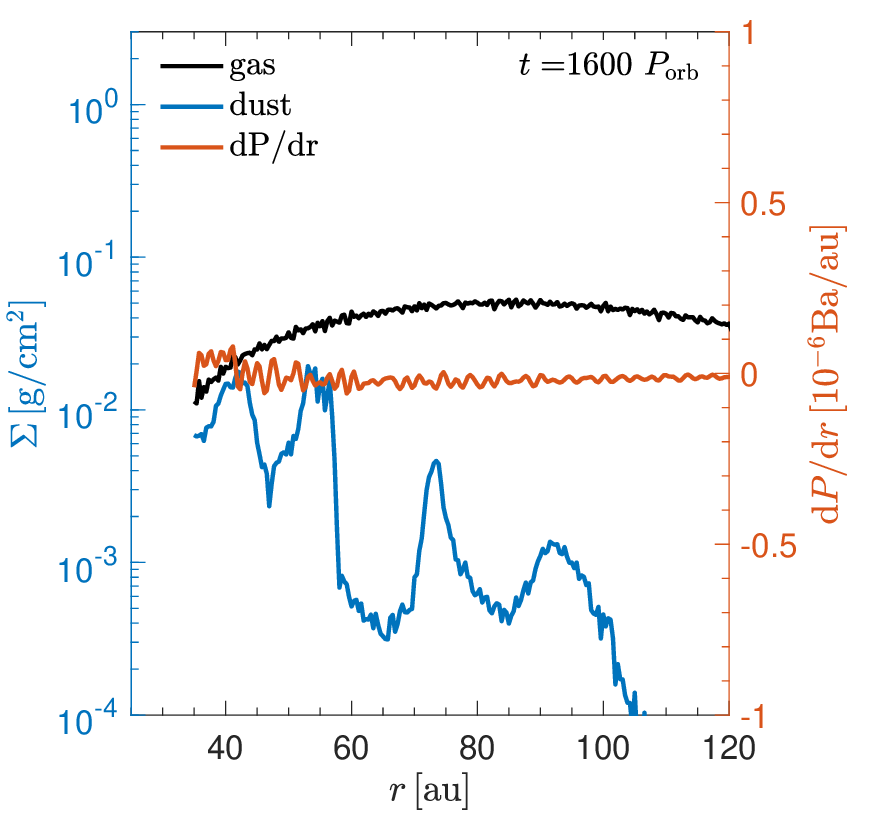}
\caption{ The gas (black) and dust (blue) surface density as a function of radius, $r$, following the left axis. The red curve denotes the pressure gradient, $\rm dP/dr$, following the right axis. The left panel shows $t = 170\, \rm P_{orb}$ and the right panel shows $t = 1600\, \rm P_{orb}$.}
\label{fig::pressure}
\end{figure*}

\section{Spiral analysis}
\label{app::spiral}

\begin{figure*} 
\centering
\includegraphics[width=0.685\columnwidth]{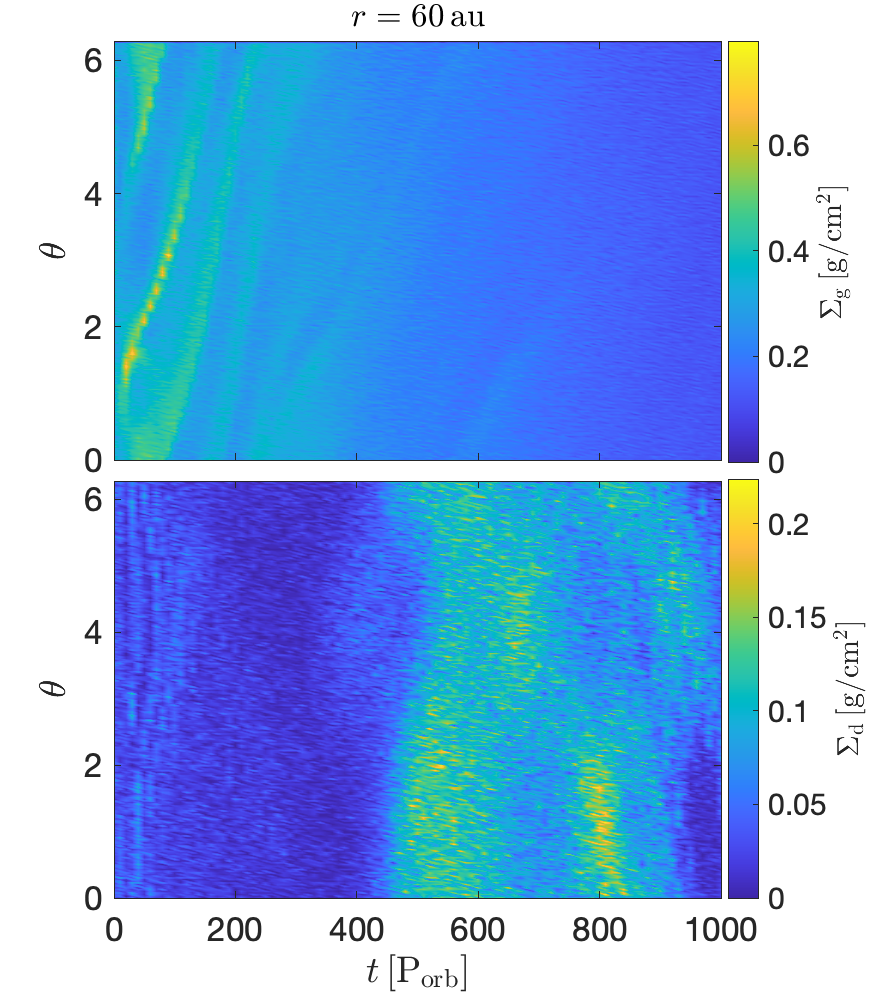}
\includegraphics[width=0.685\columnwidth]{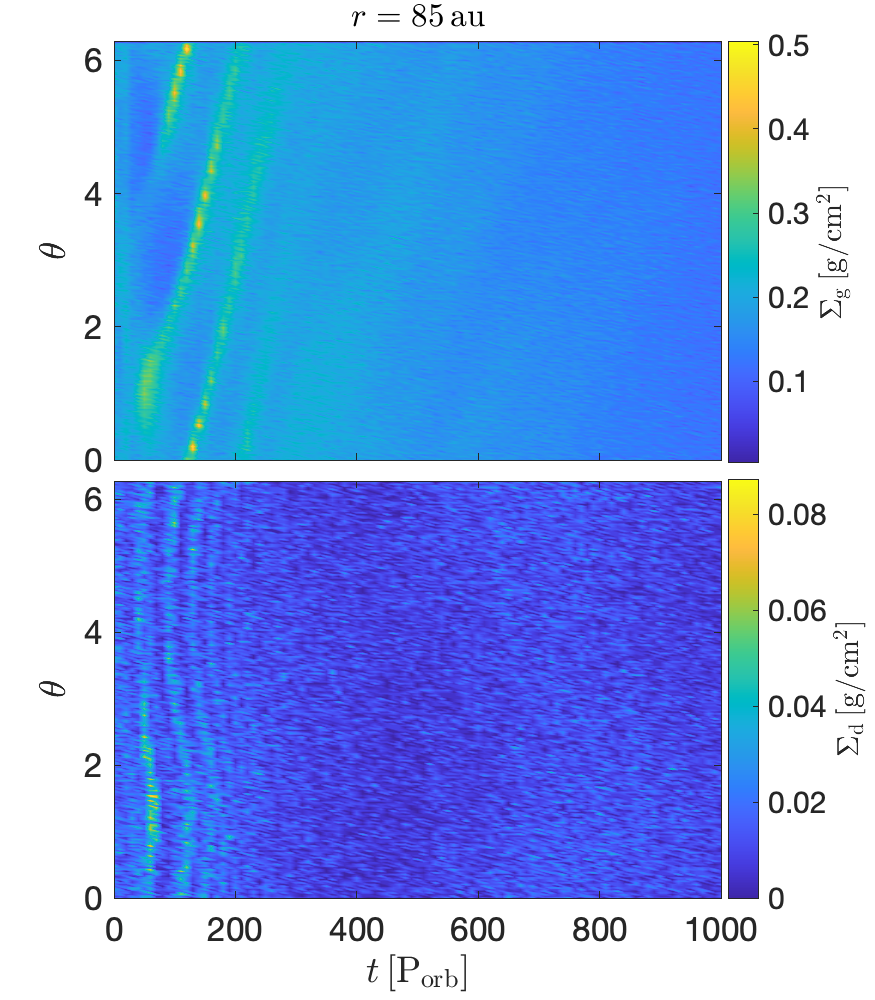}
\includegraphics[width=0.685\columnwidth]{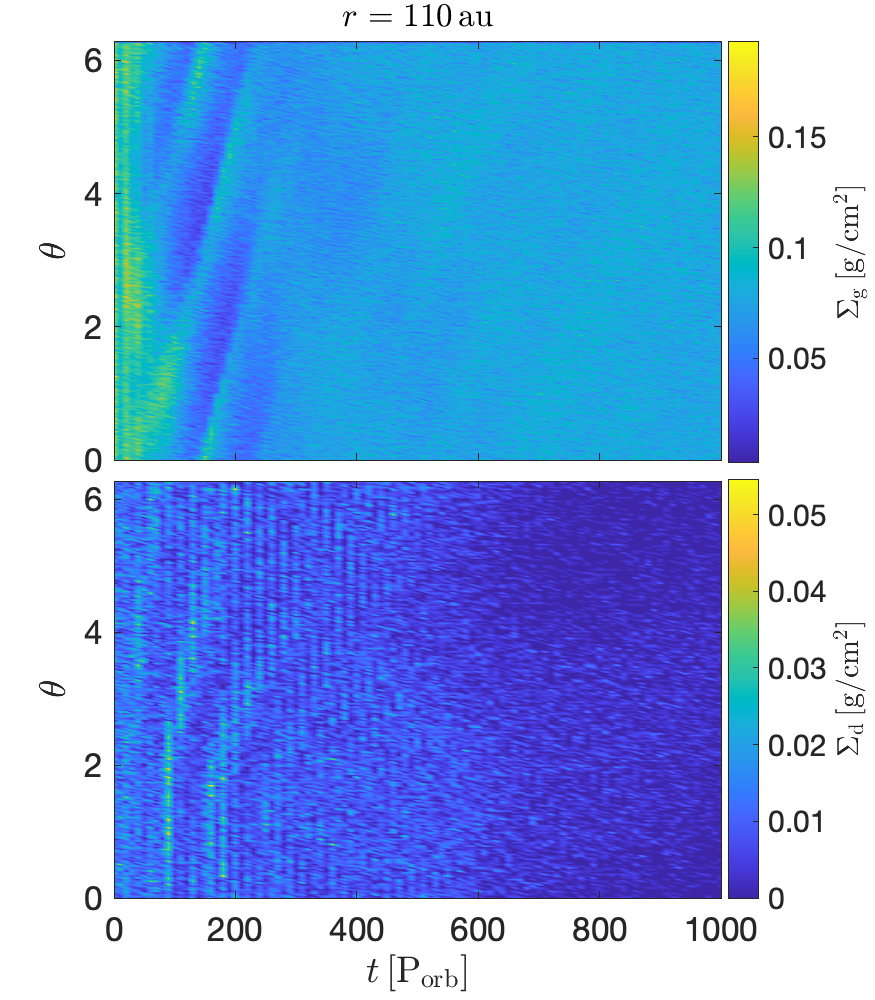}
\caption{  Anazimuthal cut of the disc surface density as a function of time in binary orbital periods, $P_{\rm orb}$. The colour maps denote the disc surface density, $\Sigma$. We analyse the azimuthal angle $\theta$ at $r = 60\, \rm au$ (left), $r = 85\, \rm au$ (middle), and $r = 110\, \rm au$ (right). The upper sub-panel represents the gas, while the lower sub-panel denotes the dust.}
\label{fig::spiral}
\end{figure*}

We now examine the spiral structure of the gas and dust. Figure~\ref{fig::spiral}  shows the azimuthal cut of the gas and dust surface densities at a radii $r = 60\, \rm au$ (left), $r = 85\, \rm au$ (middle), and $r = 110\, \rm au$ (right). There are spirals in the gas launched by the binary potential. As the radius increases, the spirals become less pronounced. At the same time, dust spirals are launched. Even though the spiral structure in the gas varies from the structure of the dusty spirals, the dust spirals are primarily launched by the binary potential. The over-dense region for $r = 60\, \rm au$ corresponds to when the initial dust traffic jams drift to $r = 60\, \rm au$. Even though the binary is launched spirals in the gas and dust, the spiral structure of the dust may also be altered by the dust traffic jams. As the gas precession deviates from that of the dust (and while they're both still tilted), the dust (especially in the outer part) crosses the gas midplane twice per orbit, producing a $m=2$ spiral structure as the dust traffic jam evolves \cite[see figures 1, 6, 7, and section 5.1 in][]{Aly2021}. 

\section{Disc morphology for h/r = 0.05}
\label{app::disc_mor}

\begin{figure*} 
\begin{center}
\includegraphics[width=1\columnwidth]{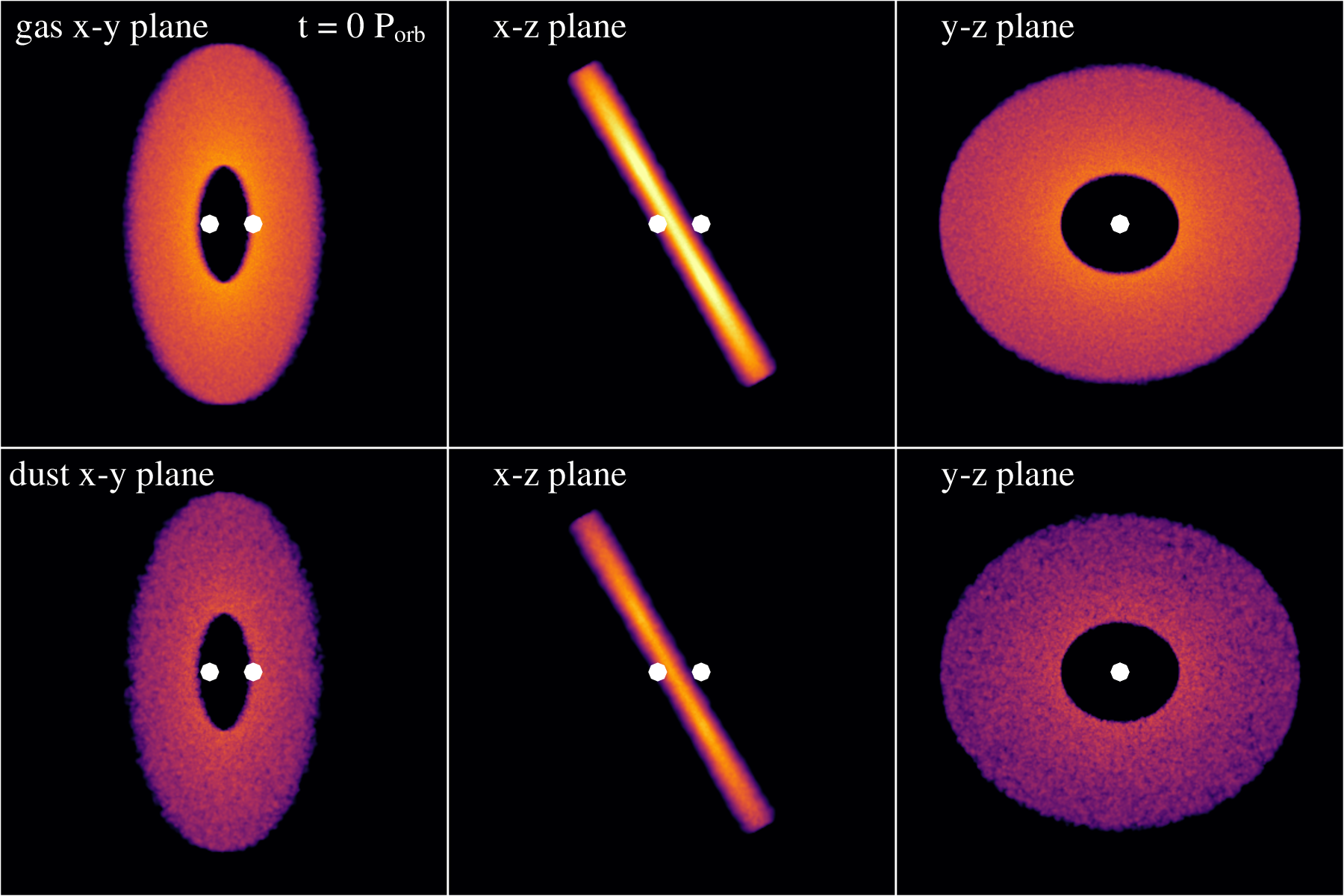}
\includegraphics[width=1\columnwidth]{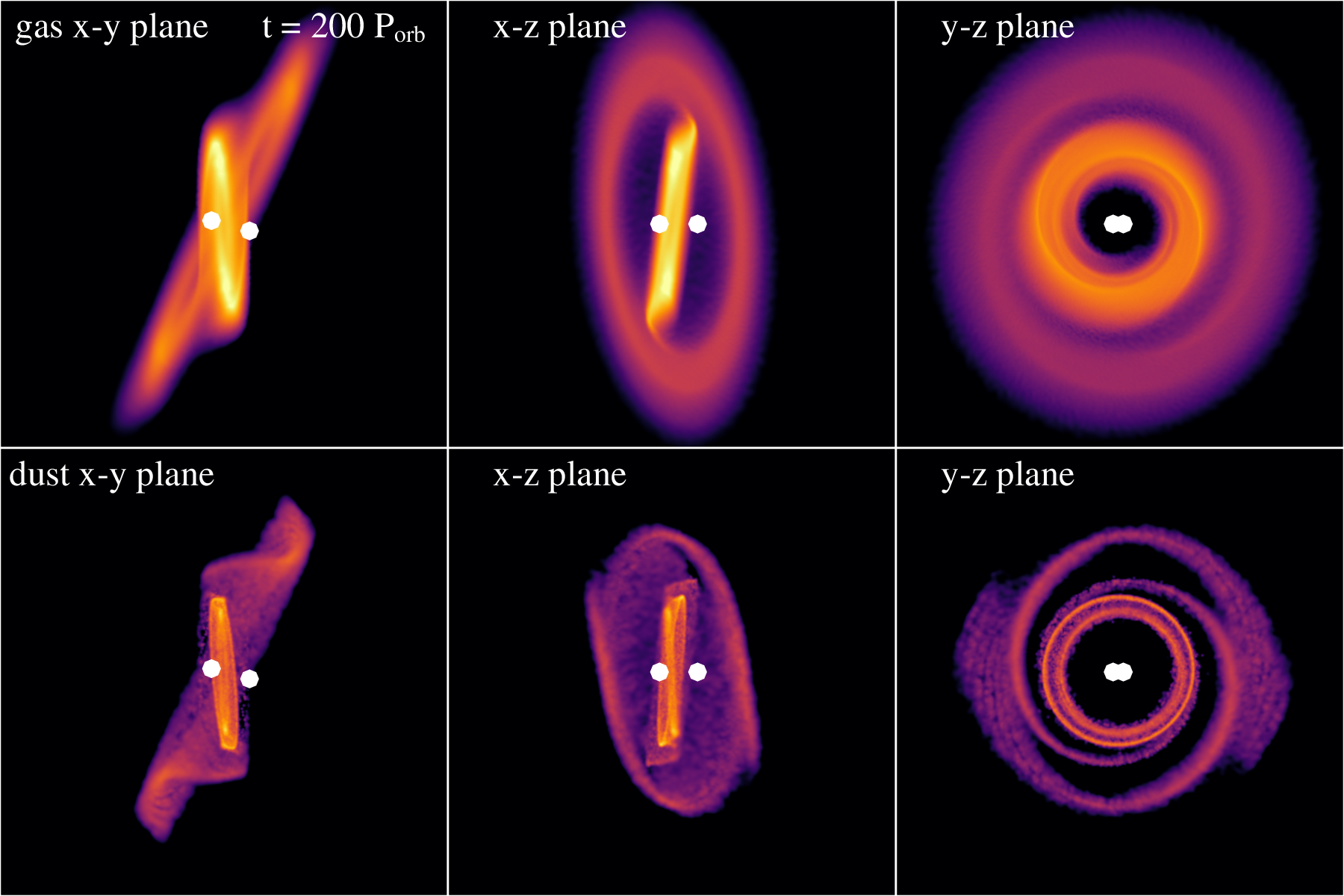}
\includegraphics[width=1\columnwidth]{plots/colorbar.eps}
\includegraphics[width=1\columnwidth]{plots/colorbar.eps}
\includegraphics[width=1\columnwidth]{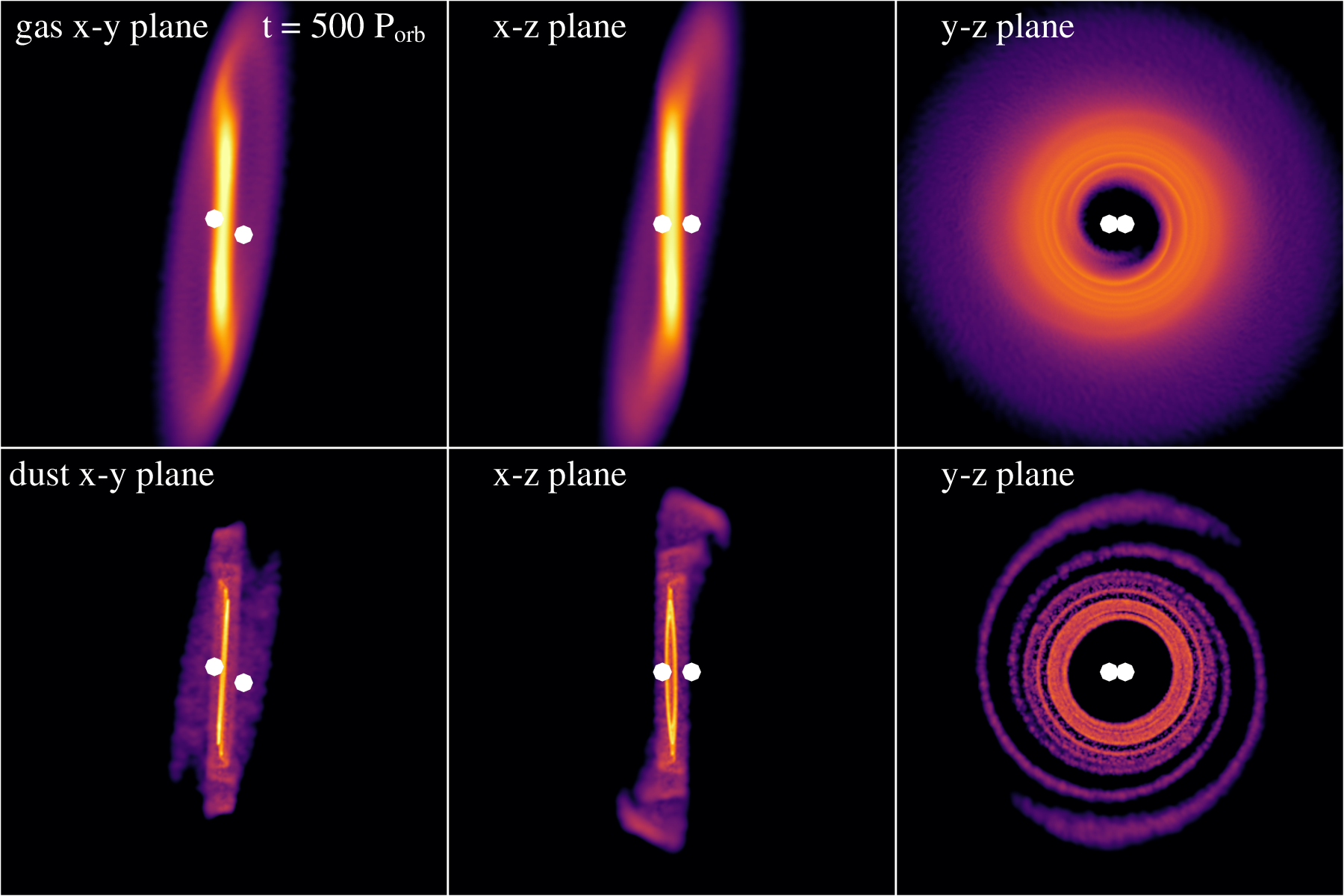}
\includegraphics[width=1\columnwidth]{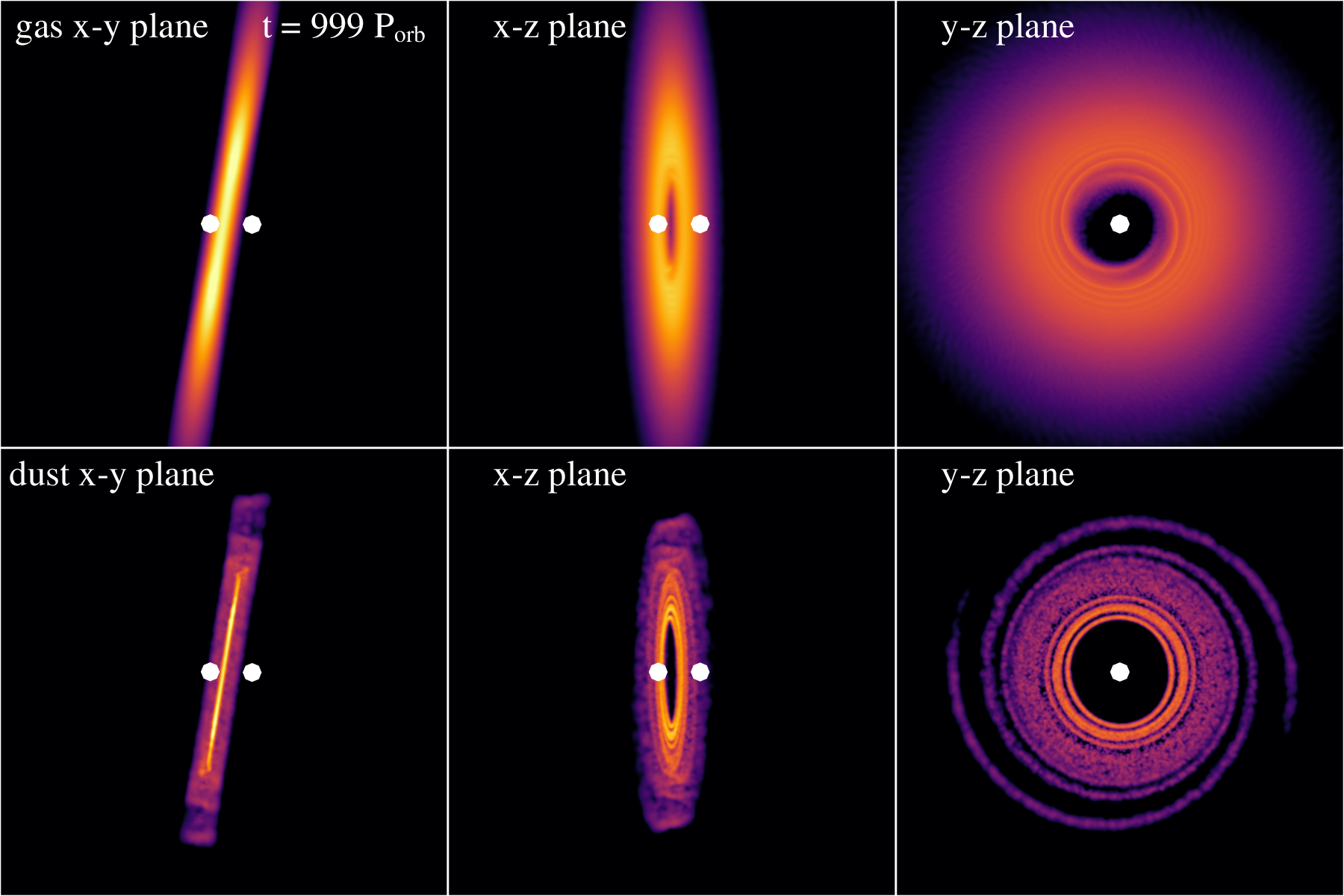}
\includegraphics[width=1\columnwidth]{plots/colorbar.eps}
\includegraphics[width=1\columnwidth]{plots/colorbar.eps}
\end{center}
\caption{ We show the circumbinary disc structure in gas and dust with $H/r=0.05$ at times $t = 0\, \rm P_{orb}$ (top-left), $t = 200\, \rm P_{orb}$ (top-right), $t = 500\, \rm P_{orb}$ (bottom-left), and $t = 1000\, \rm P_{orb}$ (bottom-right). We show the disc viewed in the $x$--$y$ plane (left sub-panel), $x$--$z$ plane (middle sub-panel), and $y$--$z$ plane (right sub-panel). The colorbar denotes the surface density.}
\label{fig::disc_mor}
\end{figure*}

\begin{figure} 
\begin{center}
\includegraphics[width=1\columnwidth]{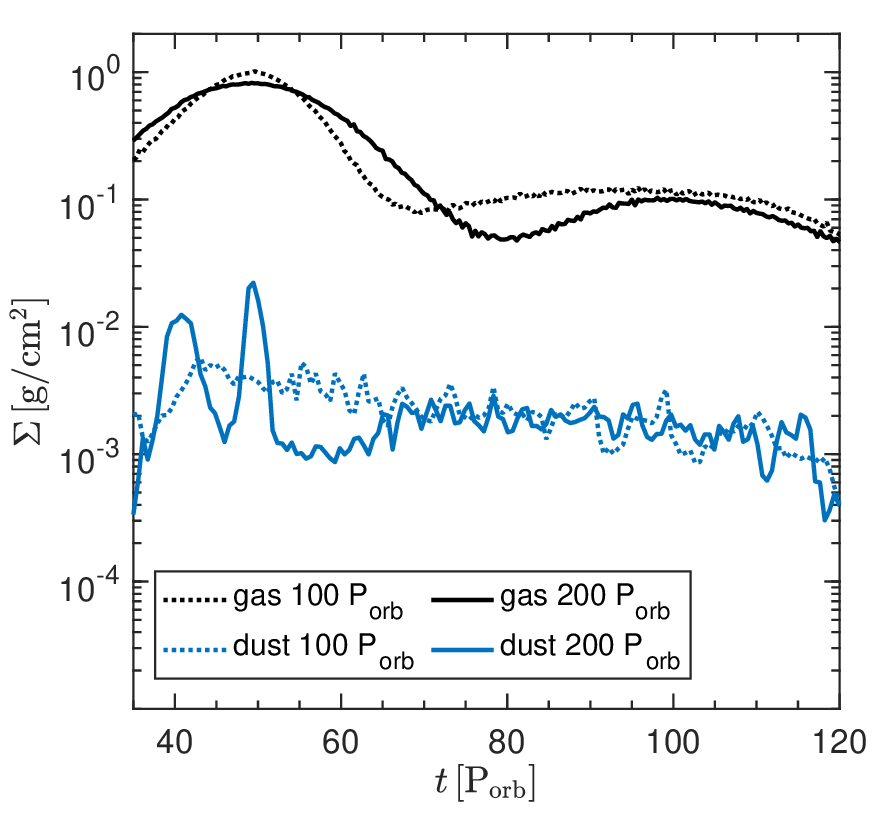}
\end{center}
\caption{ The surface density of the gas (black) and dust (blue) as a function of radius $r$ at times $t = 100\, \rm P_{\rm orb}$ (dotted) and $t = 200\, \rm P_{\rm orb}$ (solid). The gas disc breaks before the dust traffic jams form.}
\label{fig::sigma_break}
\end{figure}

 We show the disc morphology for the thin disc case, $H/r = 0.05$, in Fig.~\ref{fig::disc_mor} at times $t = 0\, \rm P_{orb}$ (top-left), $t = 200\, \rm P_{orb}$ (top-right), $t = 500\, \rm P_{orb}$ (bottom-left), and $t = 1000\, \rm P_{orb}$ (bottom-right). At $t = 200\, \rm P_{orb}$, the gas disc is broken, which causes a break in the dust. Disc breaking can complicate the dust traffic jam formation \cite[e.g.,][]{Aly2021}. Instead of a single gas disc precessing uniformly, there are now two gas discs, each with its own distinct precession frequency. The disc breaking simulations from \cite{Aly2021} show that the dust traffic jams always formed prior to the disc breaking, suggesting that the breaking was not the cause of the initial dust pileup. We conducted a similar analysis to check if the dust traffic jams formed prior to disc breaking. Figure~\ref{fig::sigma_break} shows the surface density of the gas (black) and dust (blue) as a function of radius $r$ at times $t = 100\, \rm P_{\rm orb}$ (dotted) and $t = 200\, \rm P_{\rm orb}$ (solid). At $t = 100\, \rm P_{\rm orb}$, the gas disc breaks at $r \sim 65\, \rm au$, while at the same time, there are no dust traffic jams present. At $t = 200\, \rm P_{\rm orb}$, the break in the gas disc propagates outward, with two dust traffic jams forming interior to the breaking radius. Therefore,  the gas disc breaks before the dust traffic jams form, which is opposite to the results from \cite{Aly2021}. The dust traffic jams forming within the inner disc may explain why the location of these dust traffic jams is closer versus the simulations with no disc breaking (refer back to Fig.~\ref{fig::peak}).

\bsp	
\label{lastpage}
\end{document}